\documentclass[pre,aps,onecolumn,superscriptaddress,nofootinbib]{revtex4}
\pdfoutput=1
\usepackage{amssymb}
\usepackage{epsfig}
\usepackage{amsmath}
\usepackage{subfigure}
\usepackage{graphicx}
\usepackage{textcomp}
\usepackage{url}
\usepackage{float}
\usepackage{color}
\usepackage{cases}
\usepackage[normalem]{ulem}

\usepackage[titletoc,title]{appendix}

\usepackage[colorlinks=true, urlcolor=blue, anchorcolor=blue, citecolor=blue,filecolor=blue,linkcolor=blue,menucolor=blue
]{hyperref}

\usepackage{color}

\begin{document}
\title{Geometrical optics of constrained Brownian motion: three short stories}

\author{Baruch Meerson}
\email{meerson@mail.huji.ac.il}
\author{Naftali R. Smith}
\email{naftali.smith@mail.huji.ac.il}
\affiliation{Racah Institute of Physics, Hebrew University of
Jerusalem, Jerusalem 91904, Israel}


\begin{abstract}
The optimal fluctuation method  -- essentially geometrical optics  --  gives
a deep insight into large deviations of Brownian motion. Here we illustrate this point by telling three short stories about Brownian motions, ``pushed" into a large-deviation regime by constraints. In story 1 we compute the short-time large deviation function (LDF) of the winding angle of a Brownian particle wandering around a reflecting disk in the plane.   Story 2 addresses
a stretched Brownian motion above absorbing obstacles in the plane.  We compute the short-time  LDF of the position of the surviving Brownian particle at an intermediate point. Story 3 deals with survival of a Brownian particle in 1+1 dimension against absorption by a wall which advances according to a power law $x_{\text{w}}\left(t\right)\sim t^{\gamma}$, where $\gamma>1/2$. We also calculate the LDF of the particle position at an earlier time, conditional on the survival by a later time. In all three stories we uncover singularities of the LDFs which have a simple geometric origin and can be interpreted as dynamical phase transitions.  We also use the small-deviation limit of the geometrical optics to reconstruct the distribution of \emph{typical} fluctuations. We argue that, in stories 2 and 3, this is the Ferrari-Spohn distribution.
\end{abstract}

\maketitle
\nopagebreak

\section{Introduction}

Large deviations of many stochastic systems can be accurately described by the optimal fluctuation method (OFM). This method is based on  a saddle-point evaluation of the properly constrained path integral of the stochastic system. It leads to a variational problem, where we should minimize a ``classical action" of the system over possible trajectories. When applied to Brownian motion, the OFM becomes geometrical optics \cite{GF,Ikeda2015,Holcman,Meerson2019,SmithMeerson2019}: an efficient, intuitive and easy-to-use framework for studying a whole class of systems, where additional constraints ``push" the Brownian motion into a large-deviation regime. In our recent works \cite{SmithMeerson2019,Meerson2019} we employed the geometrical optics for studying different large-deviation statistics of  Brownian excursions in $1+1$ dimension, conditioned to stay away from rapidly swinging walls. Here we illustrate the versatility and simplicity of the geometrical optics by telling three short stories about constrained Brownian motions. In two of the stories we extend this approach to $2+1$ dimensions.

Story 1 revisits the classical problem of the statistics of the winding angle of a Brownian particle, wandering around a reflecting disk in the plane \cite{Saleur,Rudnick,GF}. We focus on the short-time statistics,  which turn out to be very different from the (much better known) long-time statistics.

Story 2 was inspired by a recent work of Nechaev et al. \cite{Nechaev}. It deals with a stretched Brownian motion above an absorbing disk, and absorbing obstacles of other shapes, in the plane.  We compute the short-time  large deviation function (LDF) of the position, at an intermediate point, of the Brownian particle which has not been absorbed by the disk.
We also obtain the distribution of typical fluctuations, by mapping this model to the Ferrari-Spohn model \cite{FS} which describes a Brownian bridge in 1+1 dimension conditioned on avoiding absorption by a swinging wall. Moreover, we discuss an extension of our results to barriers of other shapes.

In story 3 we return to $1+1$ dimensions, evaluate the (exponentially small) survival probability  of a Brownian particle against absorption by a wall which advances toward the particle faster than $\sqrt{t}$, and compare our results with existing ones. In addition, we calculate the LDF of the particle position at an earlier time conditional on the survival by a later time.

In all three stories we uncover singularities of the LDFs, which have a simple geometric mechanism and can be interpreted as dynamical phase transitions of different orders. Using the \emph{small-deviation} limit of our results and additional arguments, we reconstruct the distribution of \emph{typical} fluctuations (which are normally out of reach of the OFM). Remarkably, in stories 2 and 3, the same (Ferrari-Spohn  \cite{FS}) distribution emerges in spite of the different spatial dimensions.

The starting point of our calculations is the probability of a Brownian path $\mathbf{x}\left(t\right)$, which is given, up to pre-exponential factors, by the Wiener's action, see \textit{e.g.} Ref.
\cite{legacy}:
\begin{equation}\label{Action}
-\ln P=S=\frac{1}{4D}\int_{0}^T \dot{\mathbf{x}}^2\,dt.
\end{equation}
The geometrical optics emerges from a minimization of the Wiener's action~(\ref{Action}) over trajectories $\mathbf{x}\left(t\right)$ subject to problem-specific constraints. In more than one spatial dimension, the action~(\ref{Action}) is minimized by a motion with constant speed, $|\dot{\mathbf{x}}|=\text{const}$, along the \emph{shortest} path obeying  the additional constraints. The action along such a path is
\begin{equation}
\label{eq:action_minimial_length}
S=\frac{1}{4D}\int_{0}^{T}\left(\frac{\mathcal{L}}{T}\right)^{2}\,dt=\frac{\mathcal{L}^{2}}{4DT} ,
\end{equation}
where $\mathcal{L}$ is the path's length. The problem therefore reduces to minimizing $\mathcal{L}$ under the additional constraints. Now we begin the first story.

\section{Story 1: Winding angle distribution}
\label{winding}

\begin{figure}[ht]
\includegraphics[width=0.32\textwidth,clip=]{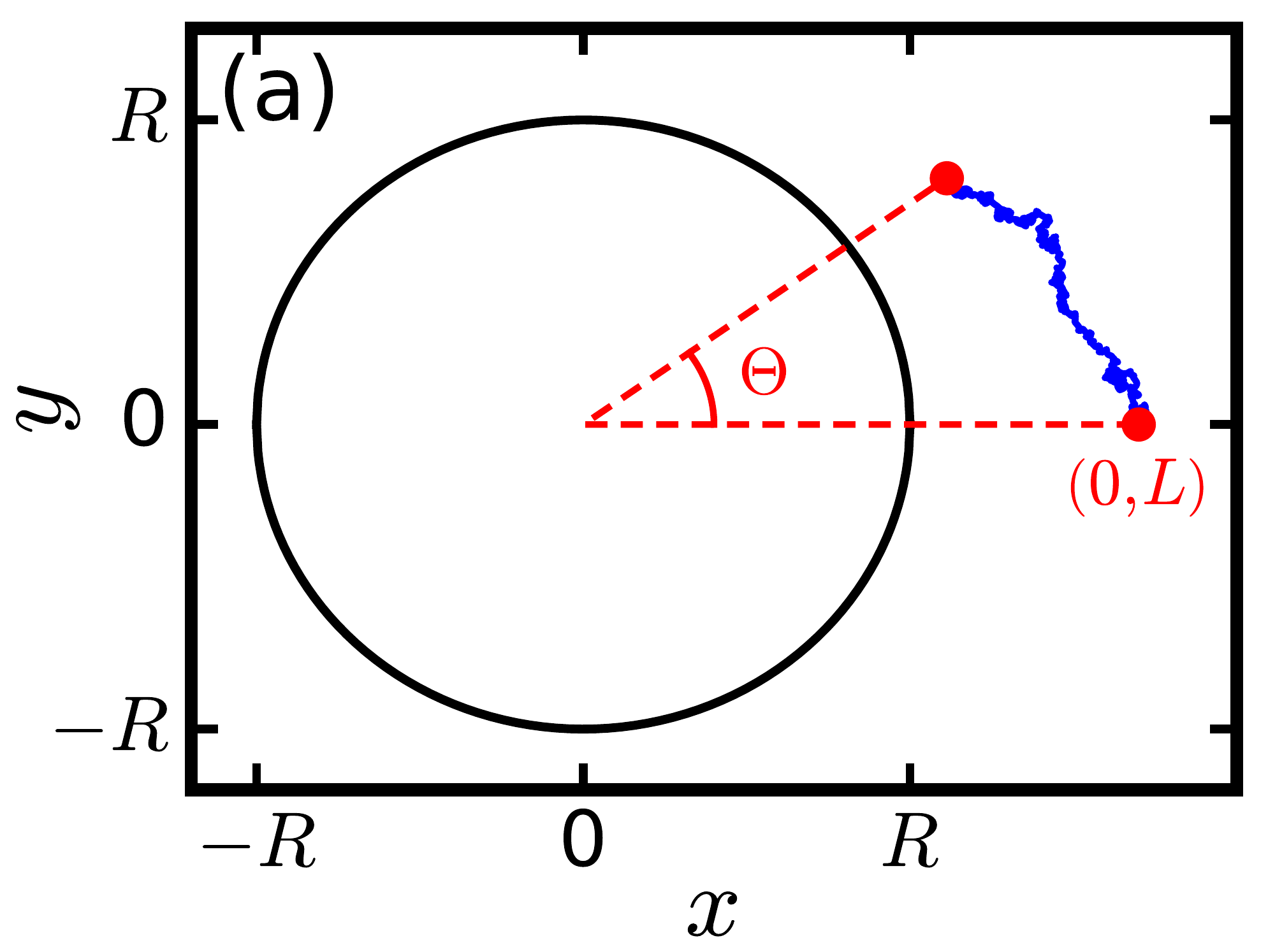}
\includegraphics[width=0.32\textwidth,clip=]{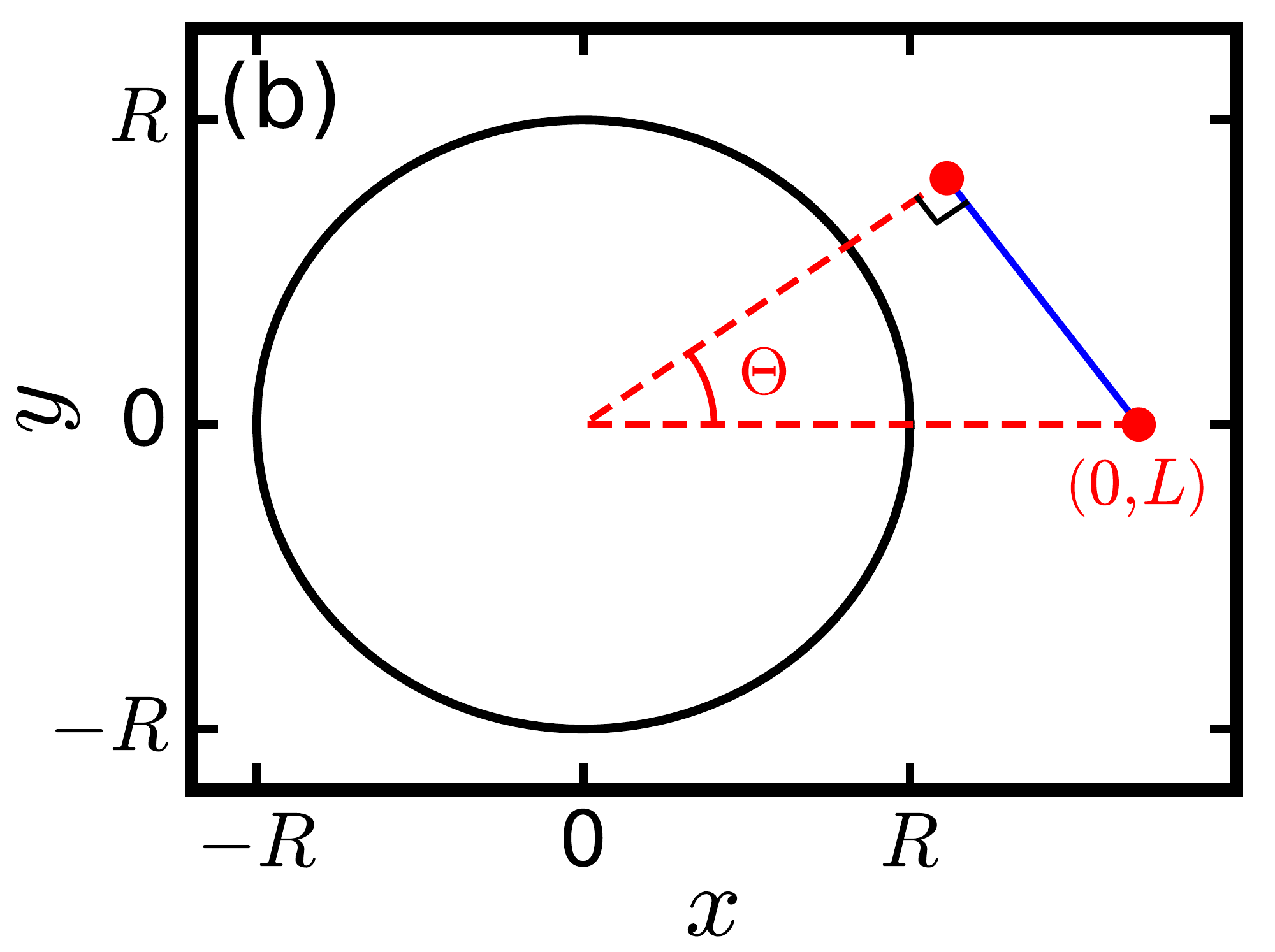}
\includegraphics[width=0.32\textwidth,clip=]{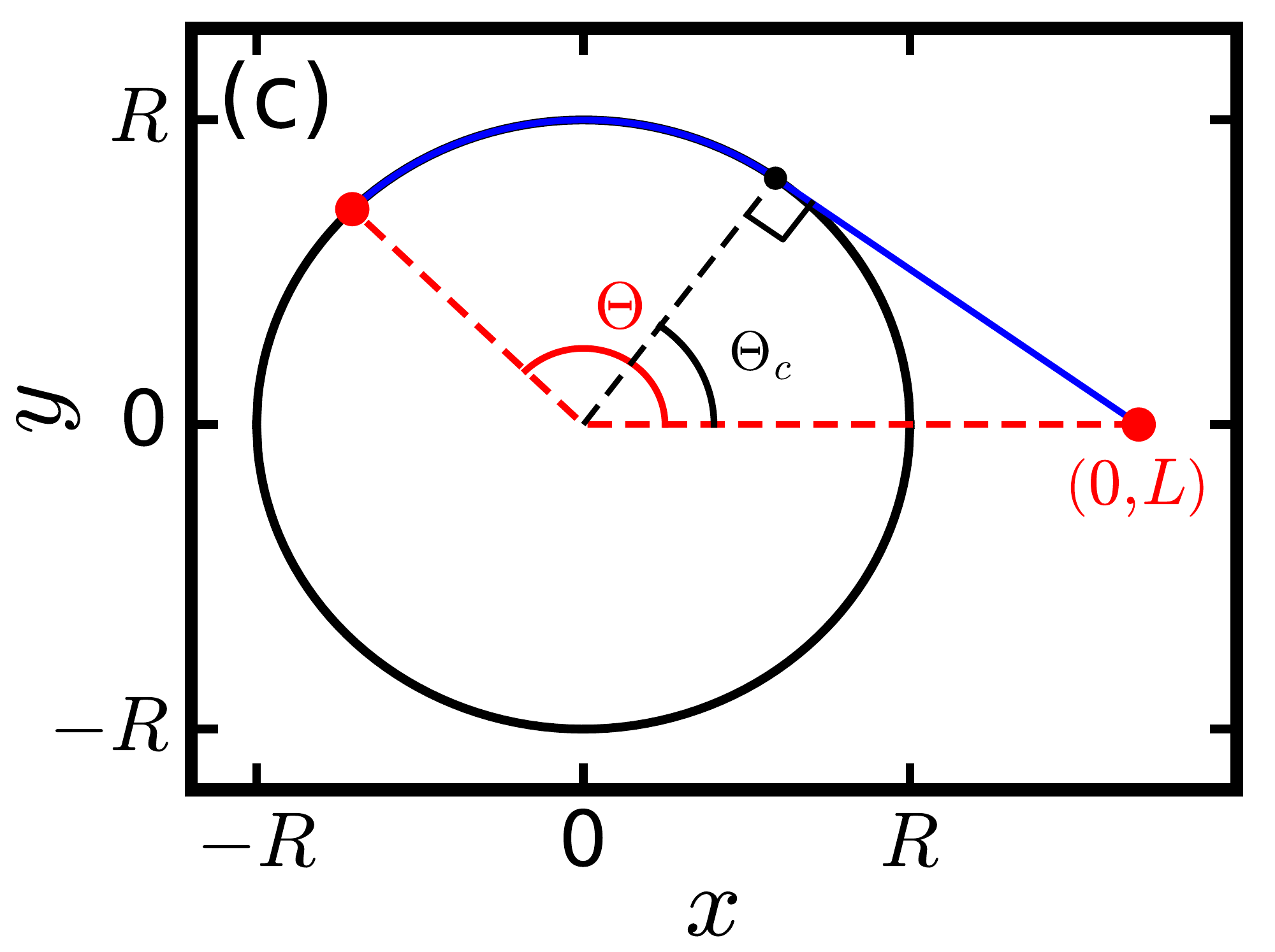}
\caption{(a) A realization of a Brownian motion in the plane outside a reflecting disk with radius $R$. We study the distribution of the winding angle $\Theta$. The optimal paths conditional on the winding angle $\Theta$ are shown in the subcritical  (b)  and supercritical (c)  regimes. At $\Theta = \Theta_c$ a second-order dynamical phase transition occurs, corresponding to a jump in the second derivative of the rate function $g\left(\Theta\right)$.}
\label{fig:winding_angle}
\end{figure}

Suppose that a Brownian particle is released at $t=0$ at a distance $L$ from the center of a reflecting disk with radius $R<L$ in the plane. 
What is the probability distribution of the winding angle $\Theta$ of the particle around the disk at time $T$, see Fig.~\ref{fig:winding_angle} (a)? This problem  was studied in Refs.~\cite{Saleur,Rudnick,GF} which focused on the long-time limit, where the characteristic diffusion length $(DT)^{1/2}$ is much larger than both $R$ and $L$, and $\Theta$ is not too large. In this limit the distribution of $\Theta$ becomes independent of $L$ and is described by the formula
\begin{equation}\label{longtime}
P\left(\Theta\right)=\frac{\pi\chi}{4\cosh^{2}\left(\pi\chi\Theta/2\right)},\quad\text{where}\quad\chi=\frac{2}{\ln\frac{4DT}{R^{2}}}.
\end{equation}
Here we are interested in the opposite, short-time limit, $(DT)^{1/2} \ll R, L-R$, where a sizable winding angle is a large deviation, and geometrical optics is perfectly suitable for its description%
\footnote{The geometrical optics was used in Ref. \cite{GF} to describe the limit of tight entanglement of a polymer around a disk.}.
As we will see shortly, the probability distribution $P(\Theta,T)$ in this limit is quite different.

\begin{figure}[ht]
\includegraphics[width=0.4\textwidth,clip=]{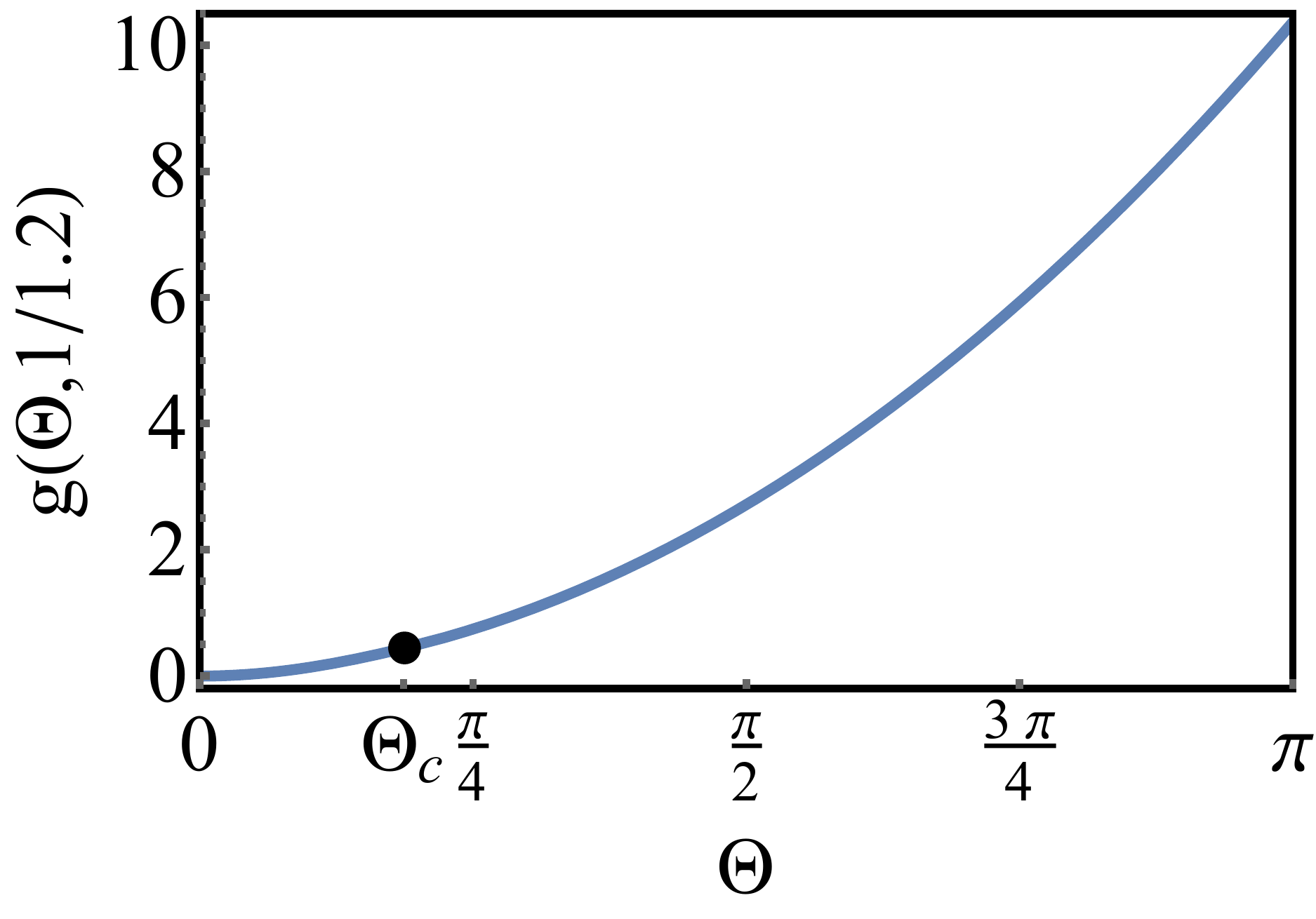}
\includegraphics[width=0.4\textwidth,clip=]{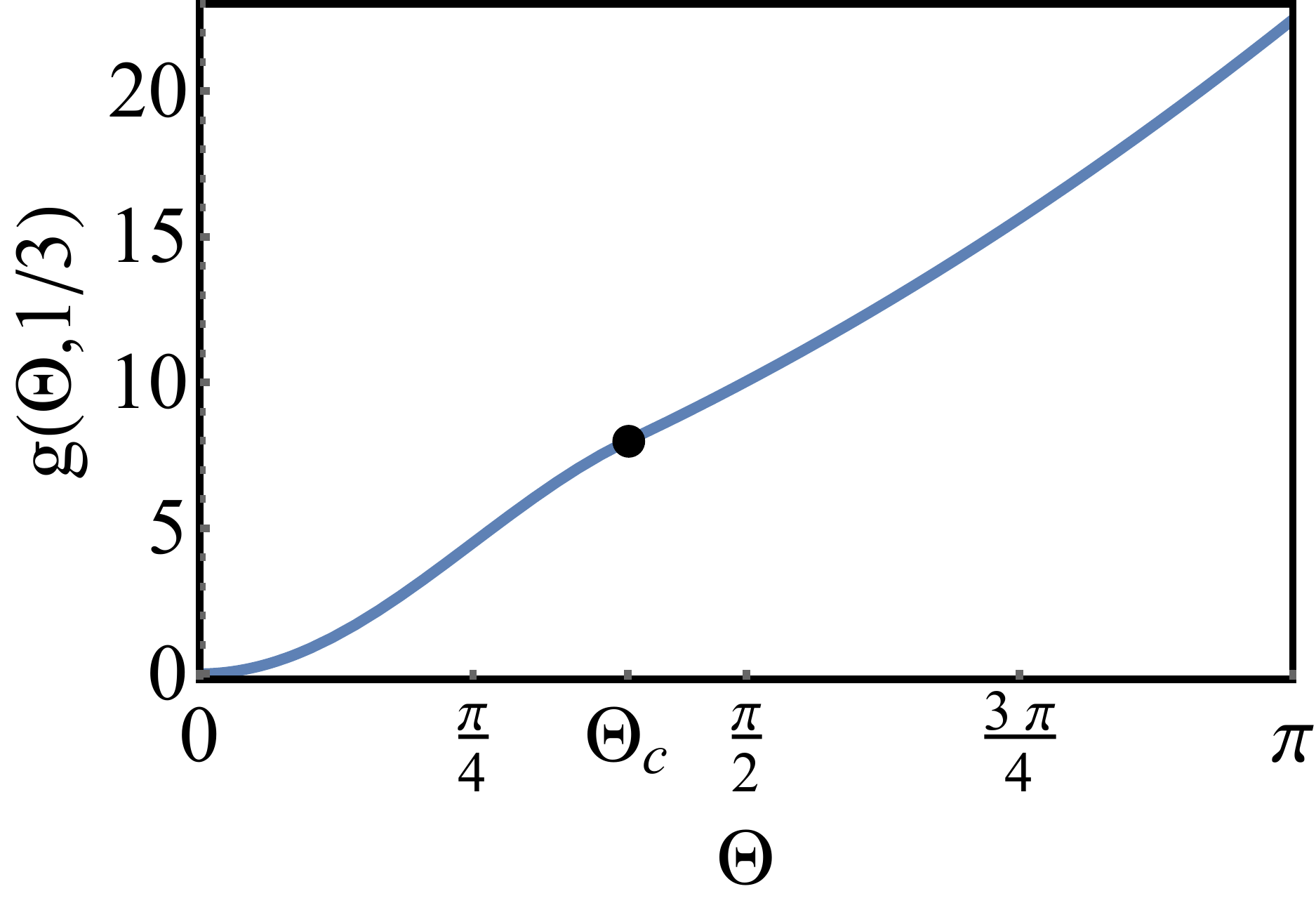}
\caption{The second-order dynamical phase transition in the winding angle distribution at short times. Shown is the rate function $g\left(\Theta,R/L\right)$, see Eq.~(\ref{anyangle}), versus the winding angle $\Theta$ for $L/R=1.2$ (left panel) and $L/R=3$ (right panel). The transition point $\Theta_{\text{c}} =\arccos\left(R/L\right)$ is indicated by the fat point. On the right panel the rate function is nonconvex on the interval $\pi/4<\Theta<\Theta_{\text{c}}$.}
\label{anglerate}
\end{figure}

In order to minimize the Wiener's action, the Brownian particle must go with a constant velocity along the shortest path. Because of the symmetry $\Theta\leftrightarrow-\Theta$ it suffices to consider $0 < \Theta < \infty$. For sufficiently small $\Theta$, the shortest path is given by the perpendicular from the initial point
$(r=L, \theta=0)$, to the ray $\theta=\Theta$, see Fig.~\ref{fig:winding_angle} (b).
The length of this path is $L \sin \Theta$, and Eqs.~(\ref{Action}) and~(\ref{eq:action_minimial_length}) yield
\begin{equation}\label{angleless}
-\ln P \simeq \frac{L^2 \sin^2 \Theta}{4D T}.
\end{equation}
This simple result, however, is valid only when $\Theta$ is less than a critical value $\Theta_{\text{c}}=\arccos\left(R/L\right)$, for which the geodesic is tangent to the disk. %
For $\Theta>\Theta_{\text{c}}$ the optimal path is given by the \emph{tangent construction} of the calculus of one-sided variations \cite{Elsgolts}. The optimal path now consists of two parts: the tangent to the disk, and the arc $\,\Theta_c<\theta<\Theta$ along the disk circumference, see Fig.~\ref{fig:winding_angle} (c).
The total length of this path is $L\sin\Theta_{\text{c}}+R\left(\Theta-\Theta_{\text{c}}\right)$,
and Eqs.~(\ref{Action}) and~(\ref{eq:action_minimial_length}) yield
\begin{equation}\label{anglemore}
-\ln P\simeq\frac{\left[L\sin\Theta_{\text{c}}+R\left(\Theta-\Theta_{\text{c}}\right)\right]^{2}}{4DT}.
\end{equation}
Overall, Eqs.~(\ref{angleless}) and (\ref{anglemore}) can be written as
\begin{equation}\label{anyangle}
-\ln P \simeq \frac{R^2}{4 DT}\,g\left(\Theta, \frac{R}{L}\right),\quad \sqrt{DT}\ll R, L-R,
\end{equation}
with the rate function
\begin{numcases}
{\!\! g(\Theta, w)=} w^{-2} \sin^2 \Theta,
& $\left|\Theta\right|\leq\arccos w$, \label{less2}\\
\left(\left|\Theta\right|+\sqrt{1/w^2-1}-\arccos w\right)^2,   & $\left|\Theta\right|\geq\arccos w$, \label{more2}
\end{numcases}
and $0<w<1$. The rate function $g\left(\Theta,w\right)$ is continuous with its first derivative with respect to $\Theta$. However, its second derivative with respect to $\Theta$ has a jump at $\Theta=\arccos w$, which can be interpreted as a second-order dynamical phase transition, see Fig.~\ref{anglerate}.  The mechanism of this transition is purely geometrical. The sharp transition appears only in the limit of $T\to 0$. It is smoothed at finite $T$,  and it  disappears at large $T$, where the distribution~(\ref{longtime}) is observed.

Notice that, for $L/R>\sqrt{2}$, the rate function $g\left(\Theta,R/L\right)$ is nonconvex, that is $\partial^{2}g/\partial\Theta^{2}<0$, for the winding angles $\pi/4<\Theta<\arccos (R/L)$, as is evident in the right panel of Fig.~\ref{anglerate}. This is one of the rare occasions when a LDF is not convex. For $1<L/R<\sqrt{2}$ the rate function is convex at all $\Theta$.

In principle,  the distribution $P\left(\Theta\right)$ can be found exactly from the solution of the diffusion equation subject to the reflecting boundary condition on the disk and a delta-function initial condition. The exact expressions for this distribution, obtained in Refs.~\cite{Rudnick,GF}, involve triple integrals of combinations of Bessel functions and trigonometric and/or exponential functions. Extracting asymptotics of these expressions is not a simple task, especially when compared with the elementary calculations we have just shown.  The authors of  Refs.~\cite{GF} succeeded in this task in the long-time limit and, in particular, arrived at the asymptotic result~(\ref{longtime}). It would be interesting to extract the short-time asymptotic~(\ref{anyangle}), including its singularities at $\Theta=\pm \Theta_{c}$, from one of the exact expression of Ref. \cite{Rudnick,GF}.

\section{Story 2: Stretched Brownian motion}

\subsection{Large deviations above a disk}

\begin{figure}[ht]
\includegraphics[width=0.4\textwidth,clip=]{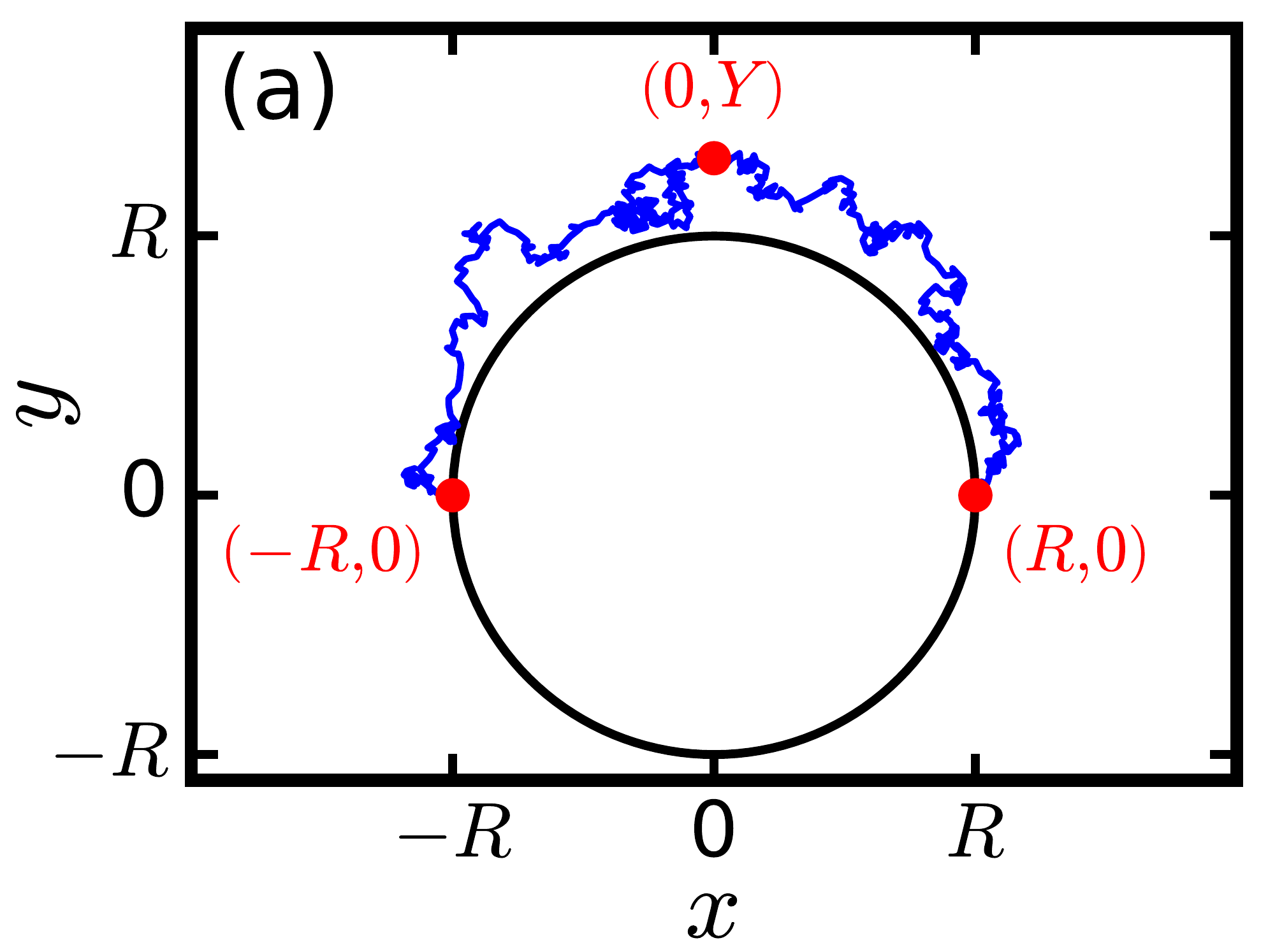}
\includegraphics[width=0.4\textwidth,clip=]{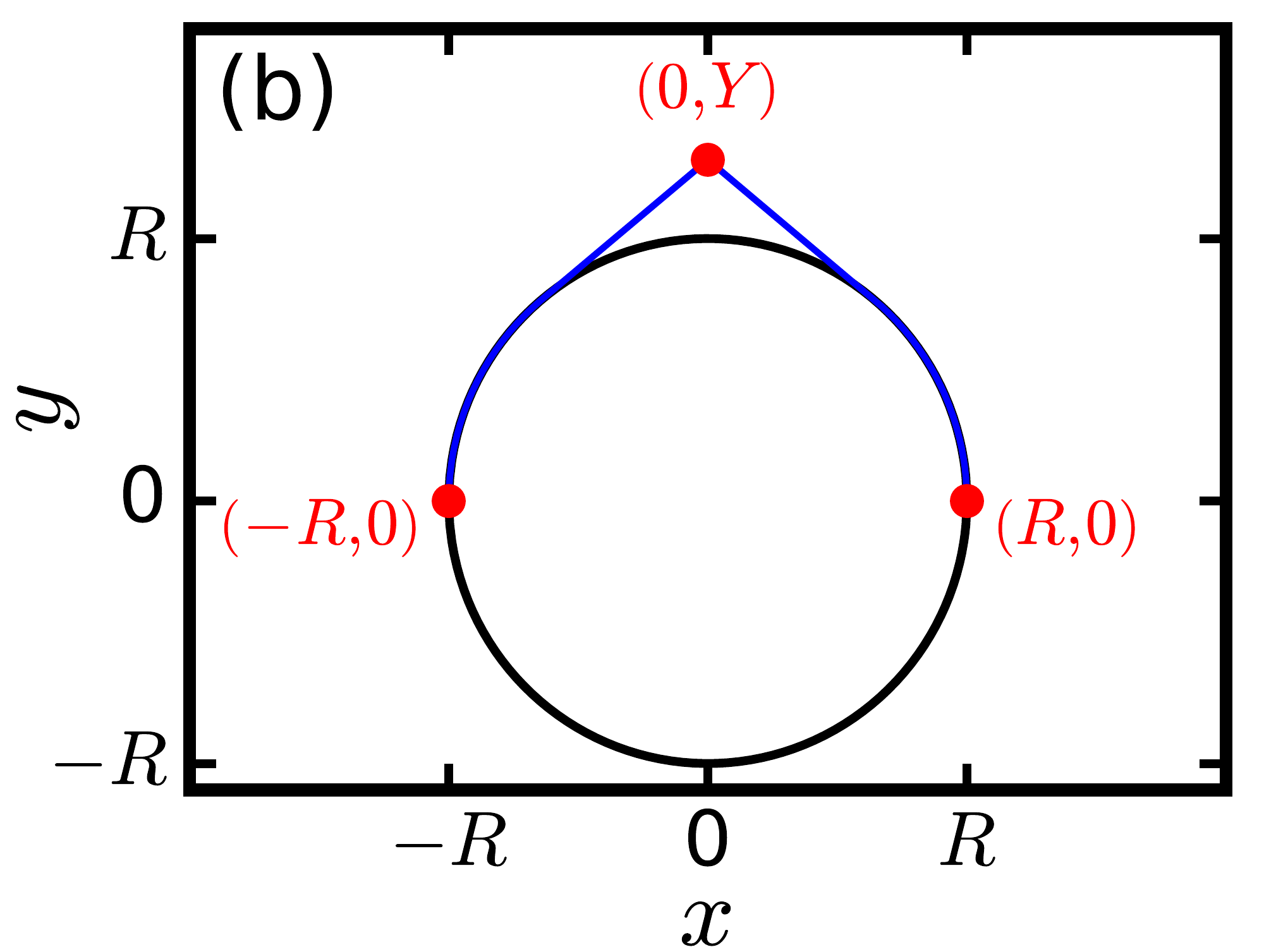}
\caption{(a) A realization of the Brownian motion in 2+1 dimensions which exits the point $\left(-R,0\right)$ and is conditional on reaching the point $\left(R,0\right)$ and avoiding absorption by a  circular wall of radius $R$. We study the distribution of $Y$, the $y$ coordinate of the Brownian motion at $x=0$.
(b) The optimal path constrained on the value of $Y$.}
\label{fig:avoiding_circle}
\end{figure}

Here we again consider a Brownian motion around a disk of radius $R$ in the plane, but this time the disk is absorbing. The Brownian particle starts at a point which is infinitesimally to the left
of
\begin{equation}
\label{eq:Nechaev_t_0}
x\left(t=0\right)=-R,\qquad y\left(t=0\right)=0
\end{equation}
and is constrained on arriving at the point
\begin{equation}
\label{eq:Nechaev_t_T}
x\left(t=T\right)=R,\qquad y\left(t=T\right)=0
\end{equation}
and avoiding being absorbed by the disk, see Fig.~\ref{fig:avoiding_circle} (a). What is the distribution
of the $y$-coordinate of the particle at $x=0$?  This question has been recently posed, in a slightly different setting%
\footnote{The setting of Ref. \cite{Nechaev} involves only the upper half plane, and the absorbing boundary also includes the two half-lines $\left|x\right|>R,\;y=0$. The difference between the two settings turns out to be inconsequential up to a normalization factor $2$.},
 by Nechaev \textit{et al}.~\citep{Nechaev}.

The distribution $P\left(Y,T\right)$ can be expressed as an infinite series, where each term involves a double integral of a combination of Bessel functions and trigonometric and/or exponential functions \citep{Nechaev}. Instead, we will evaluate $P\left(Y,T\right)$ in the short-time limit $T \ll R^2/D$, where the geometrical optics can be used
\footnote{Nechaev et al. \cite{Nechaev} imposed an additional scaling $T\sim R$. Here we do not impose any extraneous scaling of $T$ with $R$ and only assume $DT\ll R^2$.}.
In this limit, the conditional probability $P$ is given by the ratio of the probabilities of two different optimal paths: with and without the constraint $Y=y\left(x=0\right)$, and $-\ln P$ scales as%
\begin{equation}
\label{eq:probability_scaling_Nechaev_problem}
-\ln P\left(Y,T\right)\simeq\frac{R^{2}}{2DT}s\left(\frac{Y}{R}\right).
\end{equation}
For definiteness we only  solve for $Y>0$, so the optimal paths lie in the half-plane $y\ge0$. The unconstrained path simply follows the wall, and its length is $\pi R$. The constrained path includes two tangents from the point $\left(x=0,y=Y\right)$ to the circle, see Fig.~\ref{fig:avoiding_circle} (b). The total length of this path is
$$
2\left[\sqrt{Y^{2}-R^{2}}+R\arccos\left(\sqrt{\frac{Y^{2}-R^{2}}{Y^{2}}}\right)\right] .
$$
Using Eq.~(\ref{eq:action_minimial_length}), we calculate the difference of the Wiener actions evaluated on the two paths. The result yields the large-deviation function $s$ in Eq.~(\ref{eq:probability_scaling_Nechaev_problem}):
\begin{equation}
s\left(z\right)=2\left[\sqrt{z^{2}-1}+\arccos\left(\sqrt{\frac{z^{2}-1}{z^{2}}}\right)\right]^{2}-\frac{\pi^{2}}{2}.
\label{ldfcircle}
\end{equation}
The asymptotic behaviors of $s$ are
\begin{equation}
s\left(z\right)=\begin{cases}
\frac{4\sqrt{2}\,\pi}{3}\left(z-1\right)^{3/2}-\frac{3\sqrt{2}\,\pi}{5}\left(z-1\right)^{5/2}+\frac{16}{9}\left(z-1\right)^{3}+\dots, & z-1\ll1,\\
2z^{2}-\frac{\pi^{2}}{2}+2+\frac{2}{3z^{2}}+\dots, & z\gg1.
\end{cases}
\end{equation}
As a result, the near tail of the distribution, $\left(DT\right)^{2/3}\!\!/R^{1/3}\ll h\ll R\,$ (where $h=Y-R$) is a stretched exponential with power $3/2$:
\begin{equation}
\label{eq:near_tail}
P\left(h\right)\sim\exp\left(-\frac{2\sqrt{2}\,\pi R^{1/2}h^{3/2}}{3DT}\right).
\end{equation}
The characteristic decay length $\mathfrak{h}$ of the distribution scales as $\mathfrak{h} \sim (DT)^{2/3} R^{-1/3}$, that is $\mathfrak{h}/R \sim (DT/R^2)^{2/3} \ll 1$.
The correlation length $\ell_{\text{c}}$ along the $x$ axis can be estimated by evaluating the $x$ coordinates of the points of tangency on the optimal path for (the near tail of) typical fluctuations, $h\sim\mathfrak{h}$. The result is $\ell_{\text{c}}\sim \left(RDT\right)^{1/3}$, or $\ell_{\text{c}}/R \sim (DT/R^2)^{1/3} \ll 1$.   The ensuing scaling relation $\ell_{\text{c}} \sim (R \mathfrak{h})^{1/2}$ coincides with that obtained, from simple arguments, by Nechaev \textit{et~al.}~\cite{Nechaev} for $T\sim R$. Here we established it for any $DT\ll R^2$.

\subsection{Typical fluctuations above a disk and a mapping to the Ferrari-Spohn model}

We now wish to extend our results to the regime of typical fluctuations, where the geometrical optics approximation breaks down. In order to do so, we first present a different, $1+1$ dimensional model due to Ferrari and Spohn (FS) \citep{FS}.
They studied the statistics of the position, at an intermediate time $t=\tau$, of a Brownian bridge $x\left(t\right),$ when the process is constrained on staying away from an absorbing wall, that is $x\left(t\right)>x_{\text{w}}(t)$, where $x_{\text{w}}(t)$ is a semicircle, $x_{\text{w}}(t)=C\left(T^{2}-t^{2}\right)^{1/2}$. They also extended their results to other concave (that is, convex upward) functions.
FS proved that at $T \to \infty$, typical fluctuations of $\Delta X\!=\!x\left(\tau\right)-x_{\text{w}}\left(\tau\right)$ away from the moving wall obey a universal distribution which depends only on the second derivative $\ddot{x}_{\text{w}}\left(\tau\right)$.%
\footnote{\emph{Large deviations} in the FS model and in its extensions were studied in Ref.~\cite{SmithMeerson2019}.} This universal distribution can be represented as
\begin{equation}
\label{eq:FS_distribution_with_ell_and_b}
P\left(\Delta X\right)=b\,\text{Ai}^{2}\left(\ell\Delta X+a_{1}\right) ,
\end{equation}
where $\text{Ai}\left(\dots\right)$ is the Airy function, $a_1=-2.338107\dots$ is its first zero, and
$\ell=\left[-\ddot{x}_{\text{w}}\left(\tau\right)/\left(2D^{2}\right)\right]^{1/3}\!$. By using the normalization condition $\int_{0}^{\infty}P\left(\Delta X\right)d\Delta X=1$, we obtain:
\begin{equation}
 \label{eq:bsol}
b=\frac{1}{\int_{0}^{\infty}\text{Ai}^{2}\left(a_{1}+\ell \Delta X\right)d\Delta X}=\frac{\ell}{\text{Ai}'\left(a_{1}\right)^{2}},
\end{equation}
where $\text{Ai}'$ is the derivative of the Airy function with respect to its argument.
The tail $\ell \Delta X\gg 1$ of the FS distribution~(\ref{eq:FS_distribution_with_ell_and_b}) can be obtained  by taking the $z\to\infty$ asymptotic of $\text{Ai}\left(z\right)$:
\begin{equation}
\label{eq:FS_tail}
P\left(\Delta X\right)\simeq\frac{b}{4\pi\sqrt{\ell \Delta X}}\exp\left[-\frac{4}{3}\ell^{3/2}\left(\Delta X\right)^{3/2}\right].
\end{equation}
The scaling form of our near-tail result~(\ref{eq:near_tail}), a stretched exponential with power $3/2$, coincides with that of the asymptotic~(\ref{eq:FS_tail}).
This suggests that, in the regime of typical fluctuations, the two models should be related%
\footnote{Numerical evidence for the existence of such a relation already exists \cite{Nechaev}, see below.}.
Indeed, we now present an argument which establishes a formal mapping between the two models.

Even when considering, at short times $DT\ll R^{2}$, typical fluctuations $h\ll R$ in the $y$-direction, the stochastic process $\mathbf{x}(t)$ is still pushed, by the constraints~(\ref{eq:Nechaev_t_0}) and~(\ref{eq:Nechaev_t_T}), into a large-deviation regime in the $x$-direction. We can therefore approximate the stochastic particle coordinate $x\left(t\right)$ by its (deterministic!) optimal-path counterpart
\begin{equation}
\label{eq:x_deterministic}
x\left(t\right)=-R\cos\left(\frac{\pi t}{T}\right)
\end{equation}
which, in the leading order, is unaffected by $Y$.
As a result, the stochastic process $y\left(t\right)$ is effectively described by a Brownian excursion in 1+1 dimensions, constrained by the condition $y\left(t\right) > y_{\text{w}}\left(t\right)$, where
\begin{equation}
y_{\text{w}}\left(t\right)\equiv\sqrt{R^{2}-x^{2}\left(t\right)}=R\sin\left(\frac{\pi t}{T}\right)
\label{sine}
\end{equation}
is the location of an effective moving wall, which is concave.
The temporal boundary conditions are $y\left(0\right)=y\left(T\right)=0$, and we are interested in the distribution of
\begin{equation}
Y=y\left(x=0\right)=y\left(t=\frac{T}{2}\right).
\end{equation}
In this formulation the problem is identical to that of Ref.~\cite{FS}. Therefore, the distribution $P\left(h\right)$ of typical fluctuations of $h=Y-R$ coincides with the FS distribution~(\ref{eq:FS_distribution_with_ell_and_b})
with $\ell=\left[-\ddot{y}_{\text{w}}\left(T/2\right)/\left(2D^{2}\right)\right]^{1/3}\!$. For the wall function~(\ref{sine}), this is
 \begin{equation}
\label{eq:ellsol}
 \ell=\left(\frac{\pi R^{1/2}}{\sqrt{2}\,DT}\right)^{2/3},
\end{equation}
and $b=\ell/\left[2\text{Ai}'\left(a_{1}\right)^{2}\right]$ is found from the normalization condition $\int_{R}^{\infty}P\left(Y\right)dY=\int_{-\infty}^{-R}P\left(Y\right)dY=1/2$.
As to be expected, $\ell\sim1/\mathfrak{h}$.
It is remarkable that one can establish a mapping between two systems in the regime of their \emph{typical} fluctuations by using a large-deviation technique such as the OFM. The reason is the above-mentioned scale separation, guaranteed by the strong inequality $h\ll R$. This scale separation leads to the existence of a joint validity region (the near tail of the distribution) of the FS distribution and of the large-deviation tail. As a result, the distribution $P(Y,T)$ is now known for any $Y\geq 0$.

Nechaev et.~al.~\citep{Nechaev} evaluated numerically an approximate analytic expression for $P\left(Y,T\right)$ for one set of parameters%
\footnote{Nechaev et.~al.~\citep{Nechaev} assumed that the Brownian particle arrives at the point $\left(0,Y\right)$ at time $T/2$. This simplifying assumption agrees with our argument that $x(t)$ can be replaced by its deterministic counterpart~(\ref{eq:x_deterministic}), but it is an approximation. An exact numerical evaluation would use an exact expression for $P\left(Y,t_0\right)$ for arbitrary arrival time $t_0$ at the point $\left(0,Y\right)$ and averaging over all arrival times $0<t_0<T$.}.
They treated $\ell$ and $b$ as adjustable parameters and reported a very good agreement, in the region of typical fluctuations, between $P\left(Y,T\right)$ and the FS distribution~(\ref{eq:FS_distribution_with_ell_and_b}). Here we have presented an analytic argument which establishes a formal mapping between the two models and provides analytic predictions for $\ell$ and $b$. We checked our predictions for the set of parameters $D=1$, $R=100$ and $T=10^3$, used in Fig.~5 (b) of Ref.~\cite{Nechaev}. Equation~(\ref{eq:ellsol}) yields $\ell= 0.0790\dots$, which is within 2.5\% from their fitted value $0.0811$.

In the large-deviation regime, where $h$ is of order $R$ or larger, the mapping between the 2+1 dimensional model and the FS model is no longer valid. This can be seen by comparing our result~(\ref{ldfcircle}) for the 2+1 dimensional model with the corresponding large-deviation result for the FS model \citep{SmithMeerson2019}. This situation, where two models belong to the same universality class in the regime of typical fluctuations, but not for large deviations, is common. For example, typical one-point height fluctuations in the Kardar-Parisi-Zhang (KPZ) equation in 1+1 dimension, at long times and for ``droplet" initial condition, are distributed according to the same Tracy-Widom distribution that describes typical fluctuations of the largest eigenvalue of the Gaussian unitary ensemble of random matrices \citep{Spohn}. This universality, however, breaks down for large deviations. Indeed, the large deviations of height in the KPZ equation at long times \cite{LDMS2016,SMP}  are very different from those observed in the random matrices \citep{DeanMajumdar2006,DeanMajumdar2008,MajumdarVergassola,MajumdarSchehr2014}.

\subsection{Generalizing to barriers of other shapes}

Now let us go back to Eq.~(\ref{eq:near_tail}) for the near tail of the distribution $P(h)$ and generalize it by considering
a whole family of convex obstacles which behave as
\begin{equation}\label{generalobstacle}
y(x) = R - \alpha |x|^{\lambda}+\dots ,\quad \lambda>1,
\end{equation}
in a small vicinity of $x=0$. For sufficiently small $h$ the tangency points
\begin{equation}
\label{eq:xt_lambda}
x_{\text{t}}\simeq\pm\frac{h^{1/\lambda}}{\left[\alpha\left(\lambda-1\right)\right]^{1/\lambda}}
\end{equation}
are within the applicability region
of Eq.~(\ref{generalobstacle}), and we obtain
\begin{equation}
\label{eq:near_tailgeneral}
-\ln P\left(h\right)\simeq\frac{\lambda^{2}\mathcal{L}\alpha^{\frac{1}{\lambda}}h^{\frac{2\lambda-1}{\lambda}}}{\left(2\lambda-1\right)\left(\lambda-1\right)^{\frac{\lambda-1}{\lambda}}DT},
\end{equation}
where $\mathcal{L}$ is the obstacle's perimeter (which would be equal to $\pi R$ for the semi-circle). The near tail is a stretched exponential which continuously depends on $\lambda$: it changes
from an exponential tail at $\lambda=1$ (a triangular obstacle) to a Gaussian at $\lambda\to \infty$  (a locally flat obstacle).
The characteristic decay length $\mathfrak{h}$ and the correlation length $\ell_{\text{c}}$ scale as
\begin{equation}\label{decaycorrgeneral}
\mathfrak{h} \sim \left(\frac{DT}{\alpha^{\frac{1}{\lambda}}\mathcal{L}}\right)^{\frac{\lambda}{2\lambda-1}} \quad \text{and}\quad
\ell_{\text{c}}\sim  \left(\frac{DT}{\alpha^2 \mathcal{L}}\right)^{\frac{1}{2\lambda-1}}.
\end{equation}
When $\lambda$ varies from $1$ to infinity, the scaling of $\mathfrak{h}$ with $T$ varies from
$T$ to $T^{1/2}$. 
If the obstacle can be characterized by a single length scale $R$, then
\begin{equation}\label{decaycorrgeneralsimple}
\mathfrak{h} \sim \left(DTR^{-\frac{1}{\lambda}}\right)^{\frac{\lambda}{2\lambda-1}} \quad \text{and}\quad
\ell_{\text{c}} \sim \left(DT R^{2\lambda-3}\right)^{\frac{1}{2\lambda-1}}.
\end{equation}
Notice the change in the character of the $R$-dependence of $\ell_{\text{c}}$ at $\lambda=3/2$.

By analogy with the circular obstacle, we can map the 2+1 dimensional problem, in the regime of typical fluctuations, to a 1+1 dimensional model which extends the FS model by considering more general local power-law behaviors of the moving wall. The $x$-coordinate of the optimal trajectory behaves as $x\left(t\right)\simeq v\left(t-\tau\right)+\dots$ where $v=\mathcal{L}/T$ is the constant velocity along the trajectory, and $\tau$ is the time at which the optimal trajectory crosses $x=0$ [for a symmetric obstacle $y\left(x\right)=y\left(-x\right)$ this would be $\tau=T/2$]. As a result, the obstacle~(\ref{generalobstacle}) is mapped to an effective moving wall
\begin{equation}
\label{eq:gen_parabola_effective_wall}
y_{\text{w}}\left(t\right)=R-\alpha\left|\frac{\mathcal{L}\left(t-\tau\right)}{T}\right|^{\lambda}+\dots.
\end{equation}
The near tail of the FS model with such a wall was calculated in Ref.~\cite{SmithMeerson2019}, and can be used to reproduce our result~(\ref{eq:near_tailgeneral}), see Appendix. We leave the calculation of the full distribution of typical fluctuations for this family of walls to future work.

\subsection{Dynamical phase transition}

\begin{figure}[ht]
\includegraphics[width=0.32\textwidth,clip=]{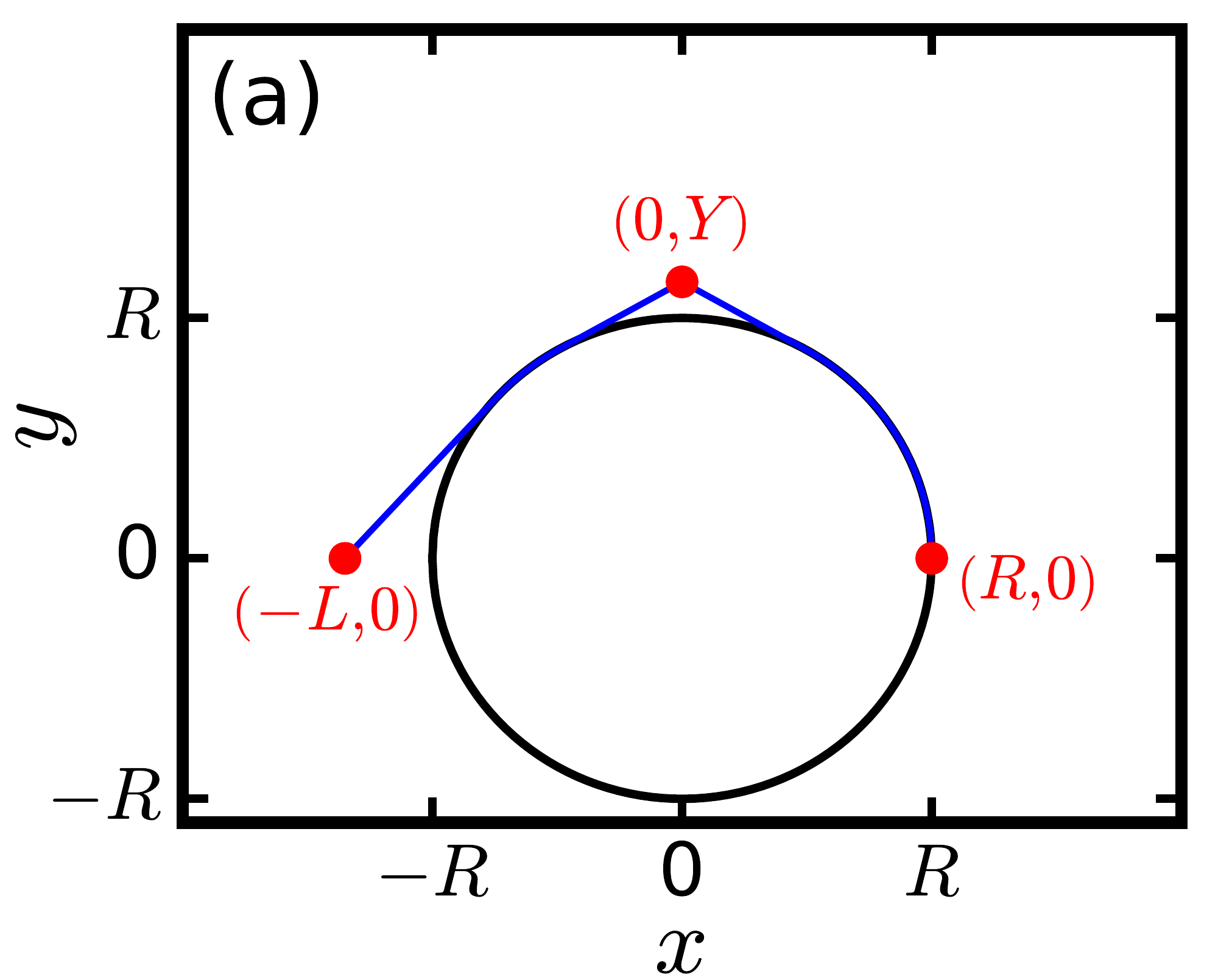}
\includegraphics[width=0.32\textwidth,clip=]{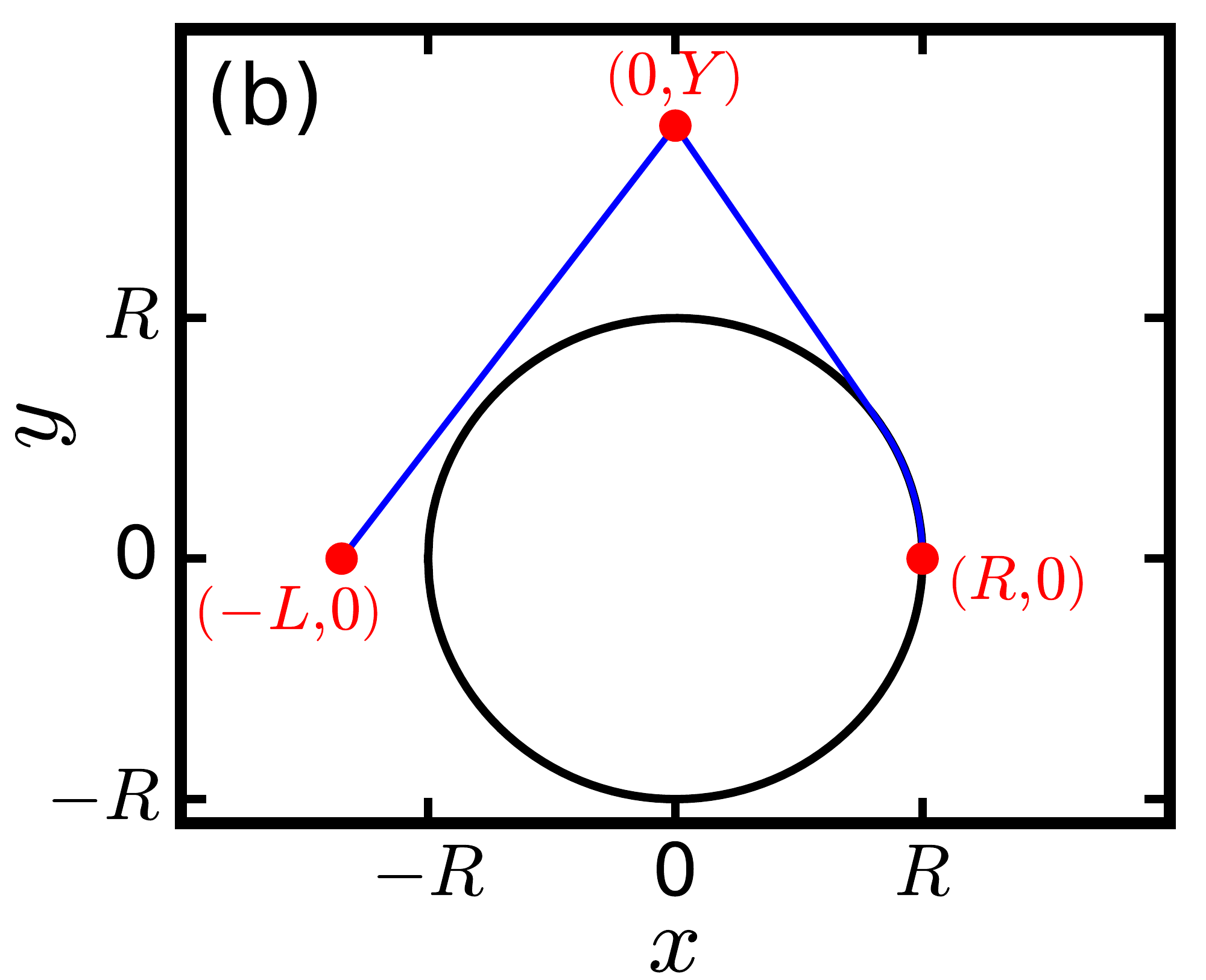}
\caption{The optimal paths for the slightly generalized problem, where the initial point is $\left(-L,0\right)$, in the  subcritical (a), and supercritical  (b) regimes. At the critical $Y$ a second-order dynamical phase transition occurs, corresponding to a jump in the second derivative of the large deviation function $s(z,L/R)$ with respect to $z$.}
\label{fig:avoiding_circle_with_L}
\end{figure}

Now let us  return to large deviations of $Y$ for the semicircular obstacle and briefly discuss a generalization of this problem, where the starting point is moved to $x\left(t=0\right)=-L,\; y\left(t=0\right)=0$.
Because of the additional length scale $L$, Eq.~\eqref{eq:probability_scaling_Nechaev_problem} gives way to
\begin{equation}
\label{eq:probability_scaling_L}
-\ln P\left(Y,T\right)\simeq\frac{R^{2}}{2DT}s\left(\frac{Y}{R},\frac{L}{R}\right).
\end{equation}
At $Y<Y_{c}\equiv RL/\sqrt{L^{2}-R^{2}}$, the optimal path is given by a tangent construction, see Fig.~\ref{fig:avoiding_circle_with_L}. However, at $Y>Y_{c}$ the  part of the optimal path, going from the point $(-L,0)$ to the point
$(0,Y)$, is just a straight line. As a result, a dynamical phase transition, corresponding to a singularity of the large deviation function $s\left(z,L/R\right)$ as a function of $z$, occurs at $z=Y_{c}/R=L/\sqrt{L^{2}-R^{2}}$. This transition is of the second order.

\section{Story 3: Particle survival against an invading wall}
\label{movingwall}

\begin{figure}[ht]
\includegraphics[width=0.32\textwidth,clip=]{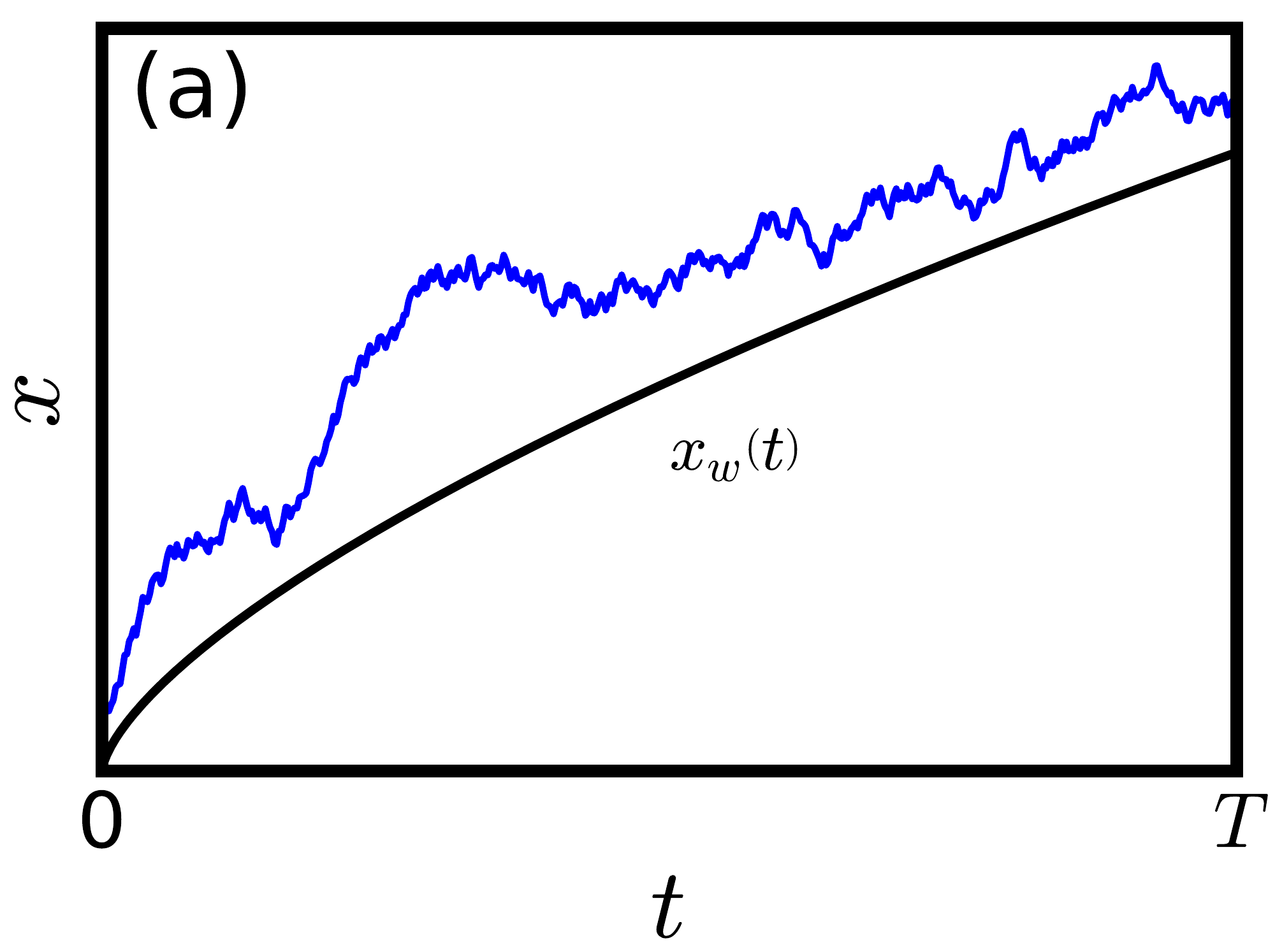}
\includegraphics[width=0.32\textwidth,clip=]{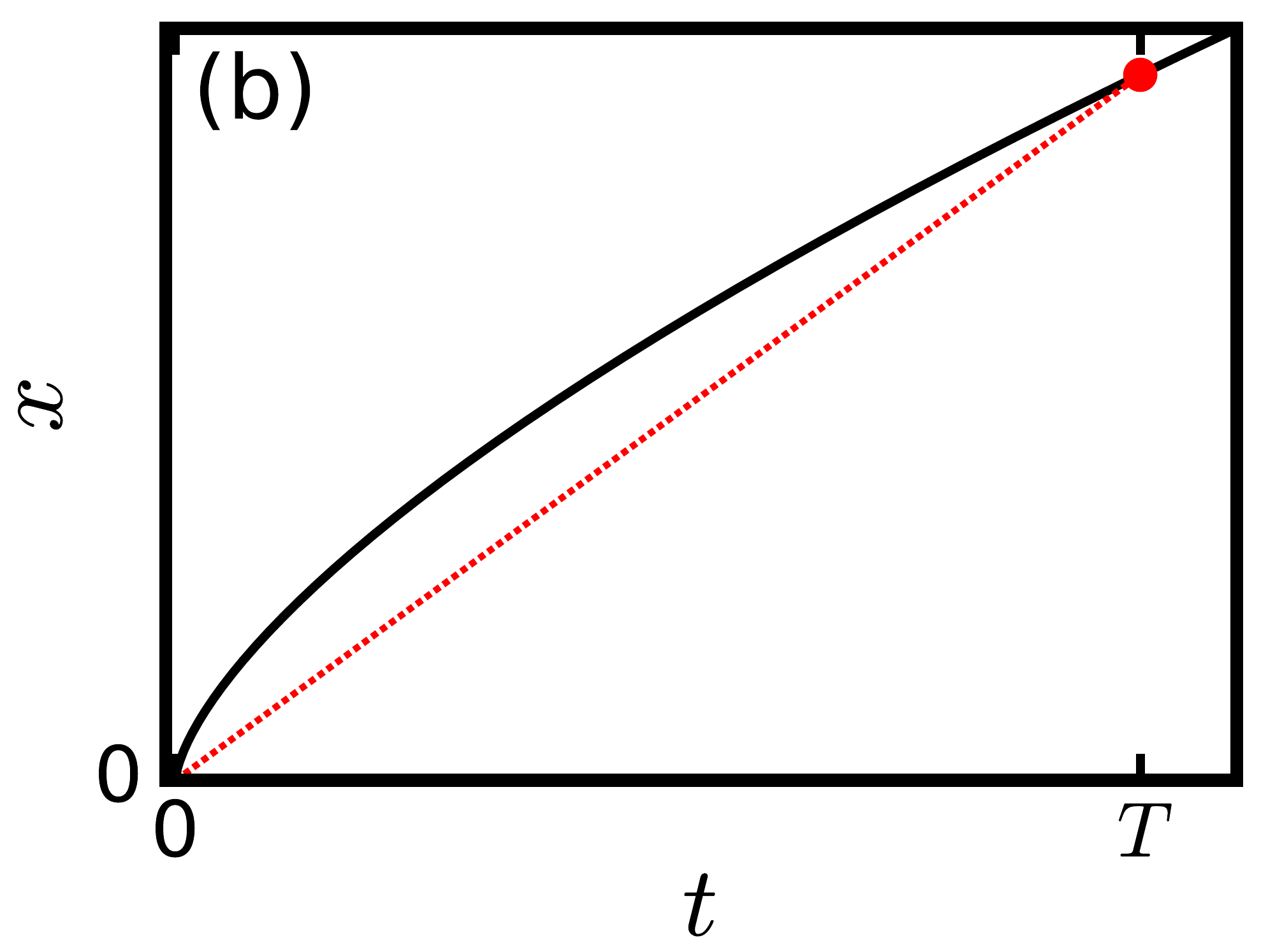}
\includegraphics[width=0.32\textwidth,clip=]{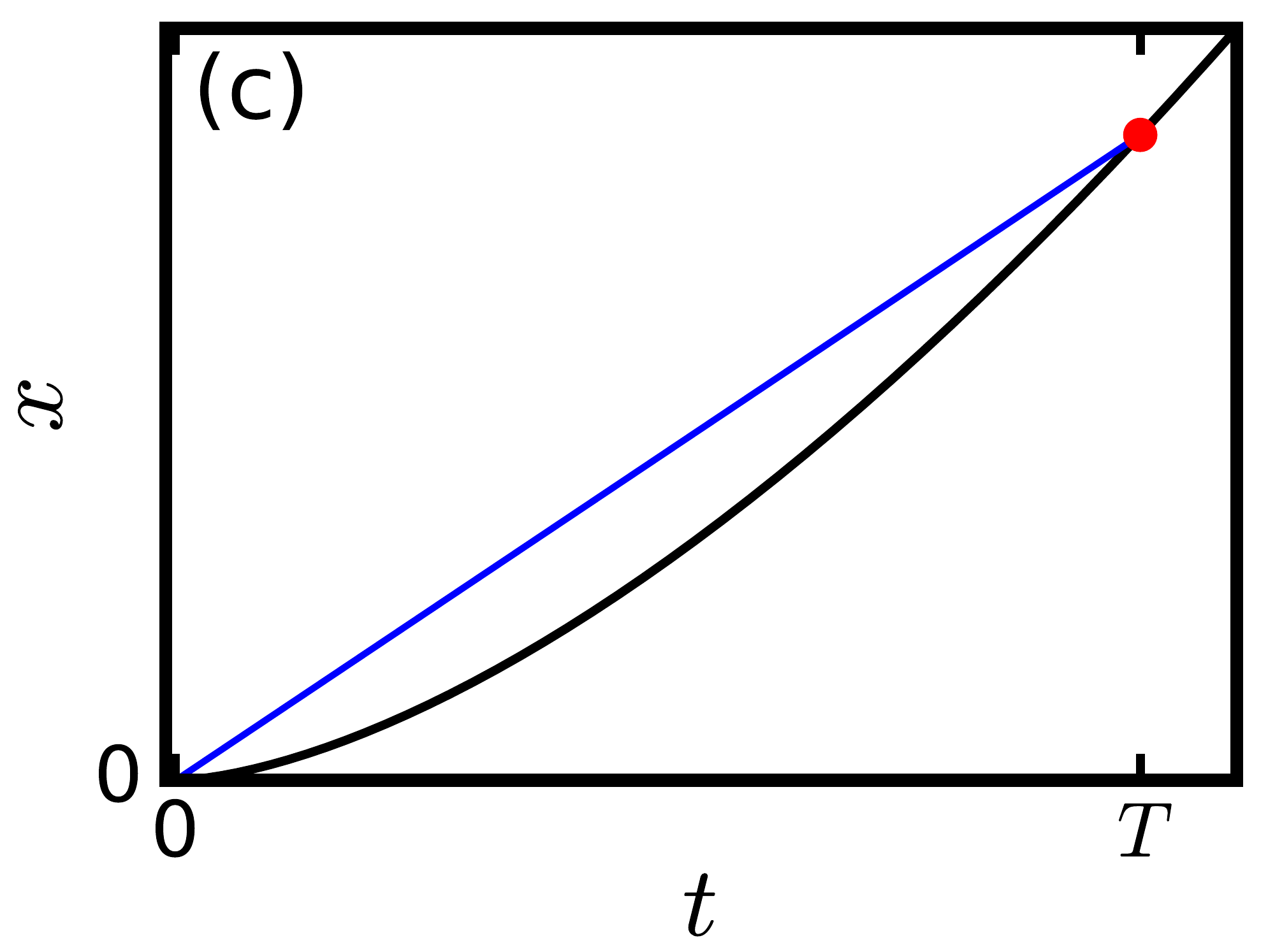}
\caption{(a) Particle survival in the presence of a moving wall $x_{\text{w}}\left(t\right)\sim t^{2/3}$. (b) The optimal path for $x_{\text{w}}\left(t\right)\sim t^{2/3}$ coincides with the  wall function, $x\left(t\right) = x_{\text{w}}\left(t\right)$. The extremal of the problem without the wall constraint (dotted) violates the constraint $x\left(t\right)> x_{\text{w}}\left(t\right)$ and is therefore not allowed. (c) The optimal path for $x_{\text{w}}\left(t\right)\sim t^{8/5}$ is of constant velocity.}
\label{fig:moving_wall}
\end{figure}

In story 3 we suppose that a Brownian particle is released at $t=0$ at $x=L>0$, whereas an absorbing wall, initially at $x=0$, is moving to the right according to a power law
\begin{equation}\label{wall}
x_{\text{w}}\left(t\right)=C t^{\gamma},
\end{equation}
where $\gamma>0$ and $C>0$ is a dimensional constant. Our first question about this system is: What is the probability that, at long time, the particle has not yet been absorbed, see Fig.~\ref{fig:moving_wall} (a)? Mathematicians dealt with this question extensively (see Ref.~\cite{Novikov1981} and references therein). The answer strongly depends on $\gamma$. For $\gamma\leq 1/2$ the survival probability decays with $T$ as a power law \cite{Novikov1981,KR1996b,RK1999,Rednerbook}.
The special case of $\gamma= 1/2$ was solved in Refs.~\cite{Sato1977,Novikov1981};  this solution was rediscovered in Refs.~\cite{KR1996b,RK1999,Rednerbook}. Here we will only be concerned with $\gamma>1/2$, when the long-time survival probability of the particle is exponentially small\footnote{The particular case $\gamma=1$ is exactly soluble for all times \cite{Rednerbook}, and we will comment on it shortly.}. The case $1/2<\gamma<1$ follows from a theorem due to Novikov \cite{Novikov1981}. Here we show how to reproduce it in a one-line calculation by using the geometrical optics. We will also consider the case $\gamma>1$.
Then we will ask an additional question about this system, which to our knowledge has not been addressed previously.

To evaluate the survival probability of the particle at $\gamma>1/2$, we should minimize the action (\ref{Action}) under a constraint that the particle stays away from the wall. At long times, $x_{\text{w}}(T)\gg L$, $L$ becomes irrelevant for the purpose of evaluating $\ln P$, so the initial condition $x\left(0\right)$ can be set to be infinitesimally close to $0$.
As in stories 1 and 2, this too is a problem of one-sided variations. For $1/2<\gamma<1$, see Fig.~\ref{fig:moving_wall} (b), the extremal of the unconstrained problem, $x\left(t\right)=x_{\text{w}}\left(T\right)t/T$, violates the constraint and therefore is not allowed. The constrained minimum is achieved when the optimal path $x\left(t\right)$ \emph{coincides} with the wall function $x_{\text{w}}\left(t\right)$, and Eq.~(\ref{Action}) yields the particle survival probability
\begin{equation}\label{Actionwall}
-\ln P\simeq \frac{1}{4D}\int_{0}^T \dot{x}_w^2(t)\,dt.
\end{equation}
In fact, Eq.~(\ref{Actionwall}) holds for a whole class of \emph{concave} (convex upwards) wall functions, like those shown in Fig.~\ref{fig:moving_wall}(b). Furthermore, Eq.~(\ref{Actionwall})  is in perfect agreement with Theorem 2 of Ref.~\cite{Novikov1981}, where it was obtained by a different, and more complicated, method.  For the power-law wall~(\ref{wall}) Eq.~(\ref{Actionwall}) yields
\begin{equation}\label{sgammaless1}
-\ln P\simeq\frac{\gamma^{2}C^{2}T^{2\gamma-1}}{4D\left(2\gamma-1\right)},\quad\frac{1}{2}<\gamma<1.
\end{equation}

For $\gamma>1$ the extremal of the unconstrained problem, $x\left(t\right)=x_{\text{w}}\left(T\right)t/T$, stays away from the wall for all $0<t<T$, see Fig.~\ref{fig:moving_wall} (c), and provides the desired minimum of the action. In this case Eq.~(\ref{Action}) yields
\begin{equation}\label{sgammamore1}
-\ln P \simeq \frac{C^2 T^{2 \gamma -1}}{4 D},\quad \gamma>1.
\end{equation}
This result can be generalized to a whole class of \emph{convex}, $\ddot{x}_{\text{w}}(t)\geq 0$, wall functions:
\begin{equation}\label{Actionpath}
-\ln P \simeq \frac{x_{\text{w}}^2(T)}{4DT}.
\end{equation}
As one can see, the probability distribution depends (up to pre-exponential factors) only on the wall position at the final time, but not on the wall history. This fact, although striking in itself, is very intuitive within the geometrical optics framework, see Fig.~\ref{fig:moving_wall} (c).

\begin{figure}[ht]
\includegraphics[width=0.4\textwidth,clip=]{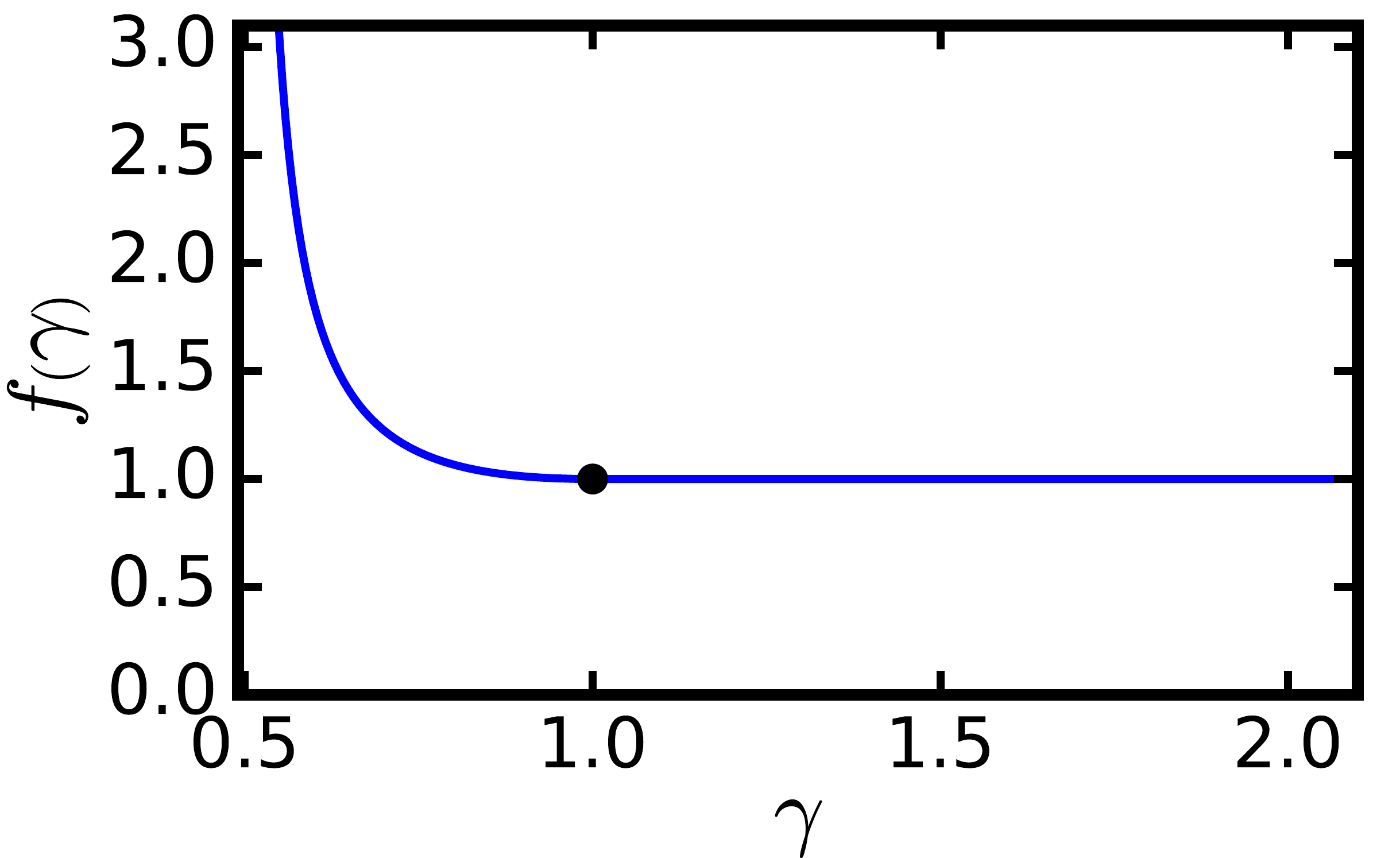}
\caption{The function $f(\gamma)$ which describes the survival probability, see Eq.~(\ref{sanygamma}).}
\label{fig:moving_wall_ldf}
\end{figure}

Equations~(\ref{sgammaless1}) and (\ref{sgammamore1}) can be written as
\begin{equation}\label{sanygamma}
-\ln P \simeq\frac{C^2 f(\gamma)T^{2 \gamma -1}}{4 D}, \quad \text{where} \quad f\left(\gamma\right)=\begin{cases}
\frac{\gamma^{2}}{2\gamma-1}, & 1/2<\gamma\leq 1,\\
1, & \gamma\geq 1.
\end{cases}
\end{equation}
In contrast to stories 1 and 2, where the geometrical optics gave short-time asymptotics of the desired statistics, Eq.~(\ref{sanygamma}) is accurate at long times. As we see, for all $\gamma>1/2$ the survival probability~(\ref{sanygamma}) is described by a stretched exponential of time with the power $2\gamma-1$\footnote{The power $2\gamma-1$ was predicted in Ref. \cite{Rednerbook}, see Eq. (4.8.1) there.}, but the function $f(\gamma)$ is non-analytic, see Fig.~\ref{fig:moving_wall_ldf}. It is continuous together with its first derivative, but its second derivative has a jump at $\gamma=1$, that is when the absorbing wall moves with a constant speed $C$.
The latter case is soluble exactly for any $T>0$ \cite{Rednerbook}. The long-time asymptotic of the exact survival probability is \cite{Rednerbook,Rednererrata}
\begin{equation}\label{gamma1exact}
P\simeq \sqrt{\frac{4}{\pi}}\, \frac{L}{\sqrt{DT}}\,\frac{D}{C^2 T} \,e^{-\frac{C^2 T}{4D}}.
\end{equation}
The geometrical-optics result (\ref{sanygamma}) for $\gamma=1$ agrees with Eq.~(\ref{gamma1exact}) up to pre-exponential factors in~(\ref{gamma1exact}). This example is instructive, as it shows an intrinsic limitation of the geometrical optics: the  pre-exponential factors $L/\sqrt{DT}$ and $D/(C^2T)$ are both very small (and therefore interesting), but they are missed by the geometrical optics.

\begin{figure}[ht]
\includegraphics[width=0.32\textwidth,clip=]{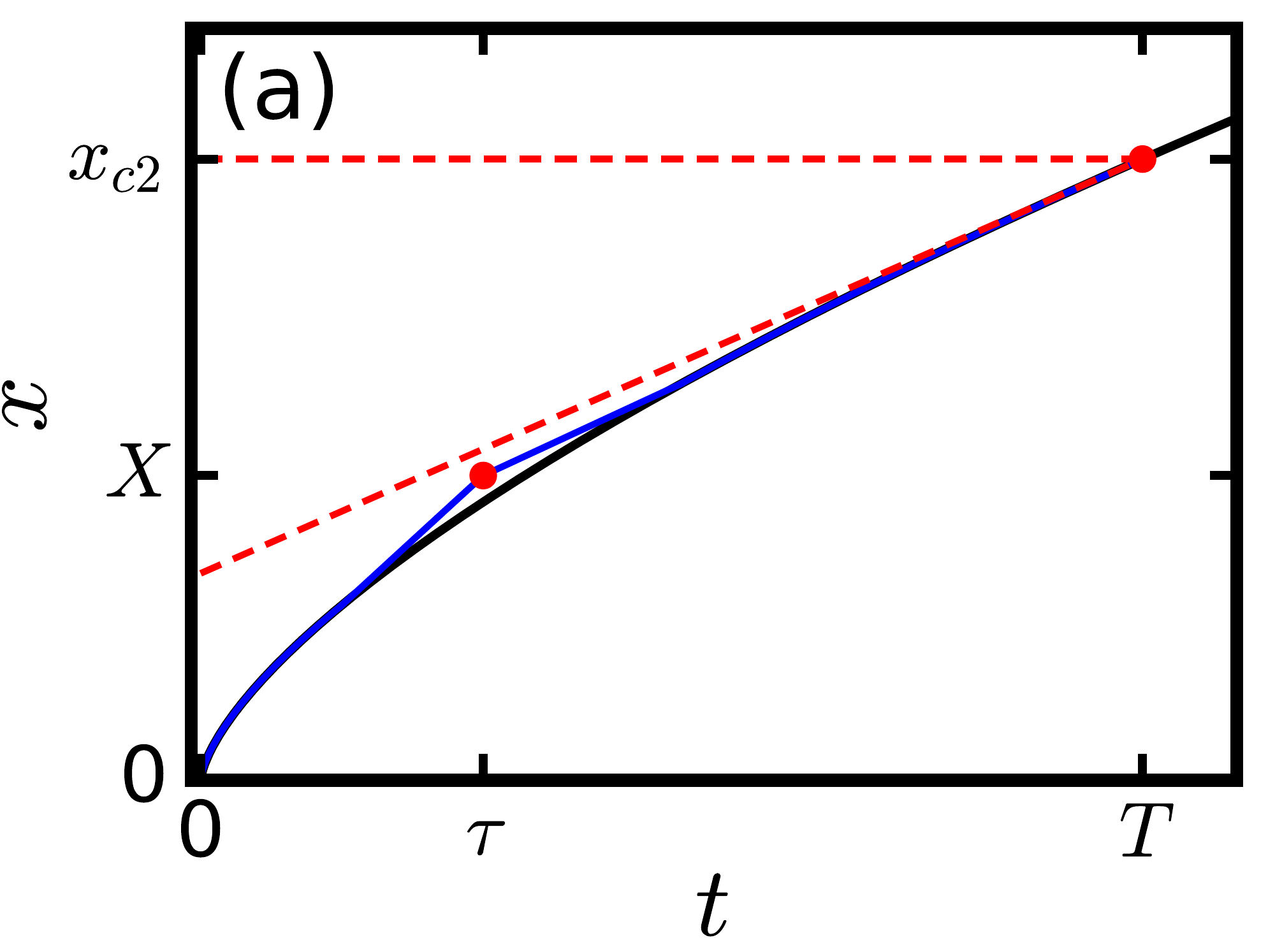}
\includegraphics[width=0.32\textwidth,clip=]{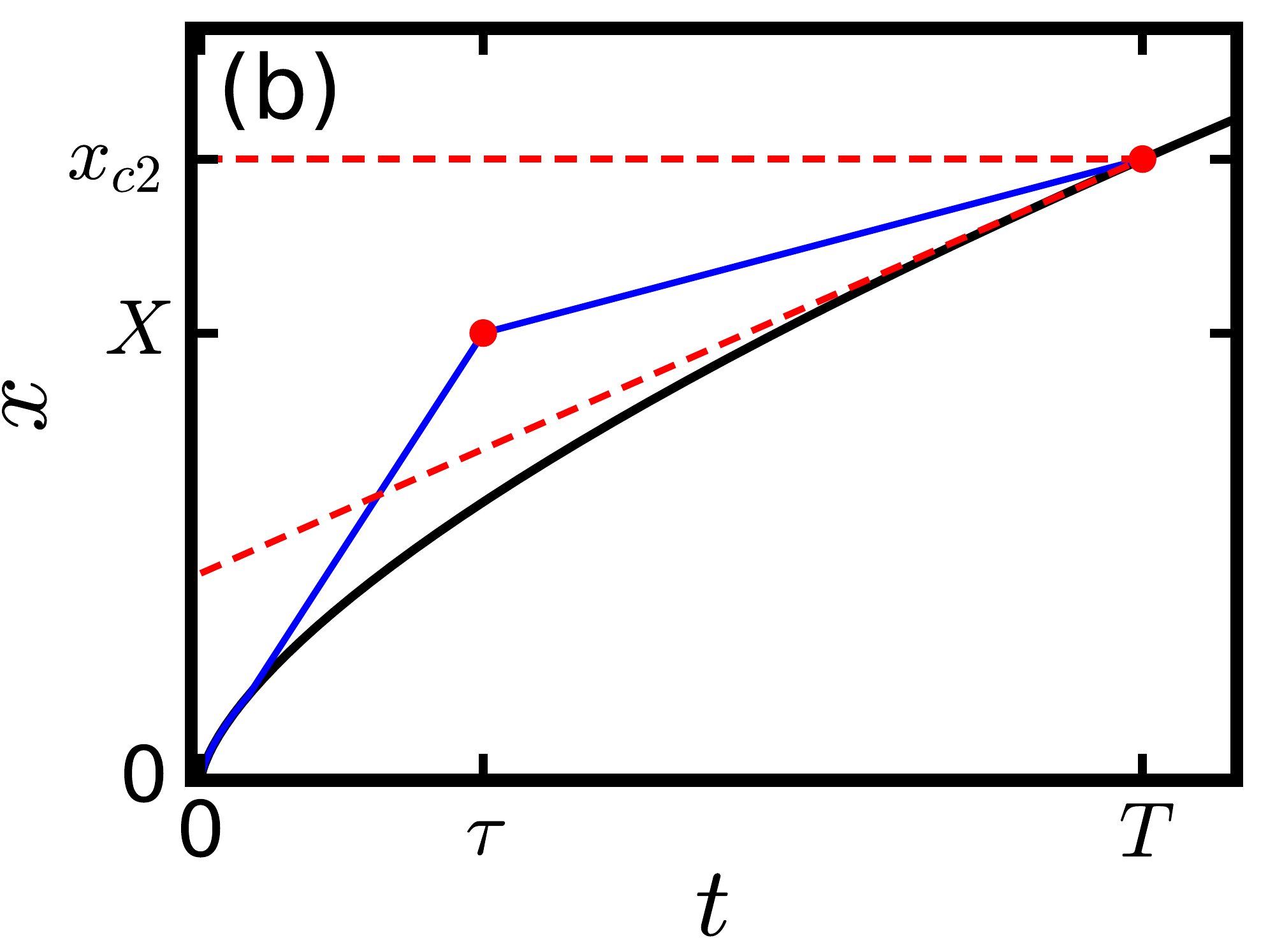}
\includegraphics[width=0.32\textwidth,clip=]{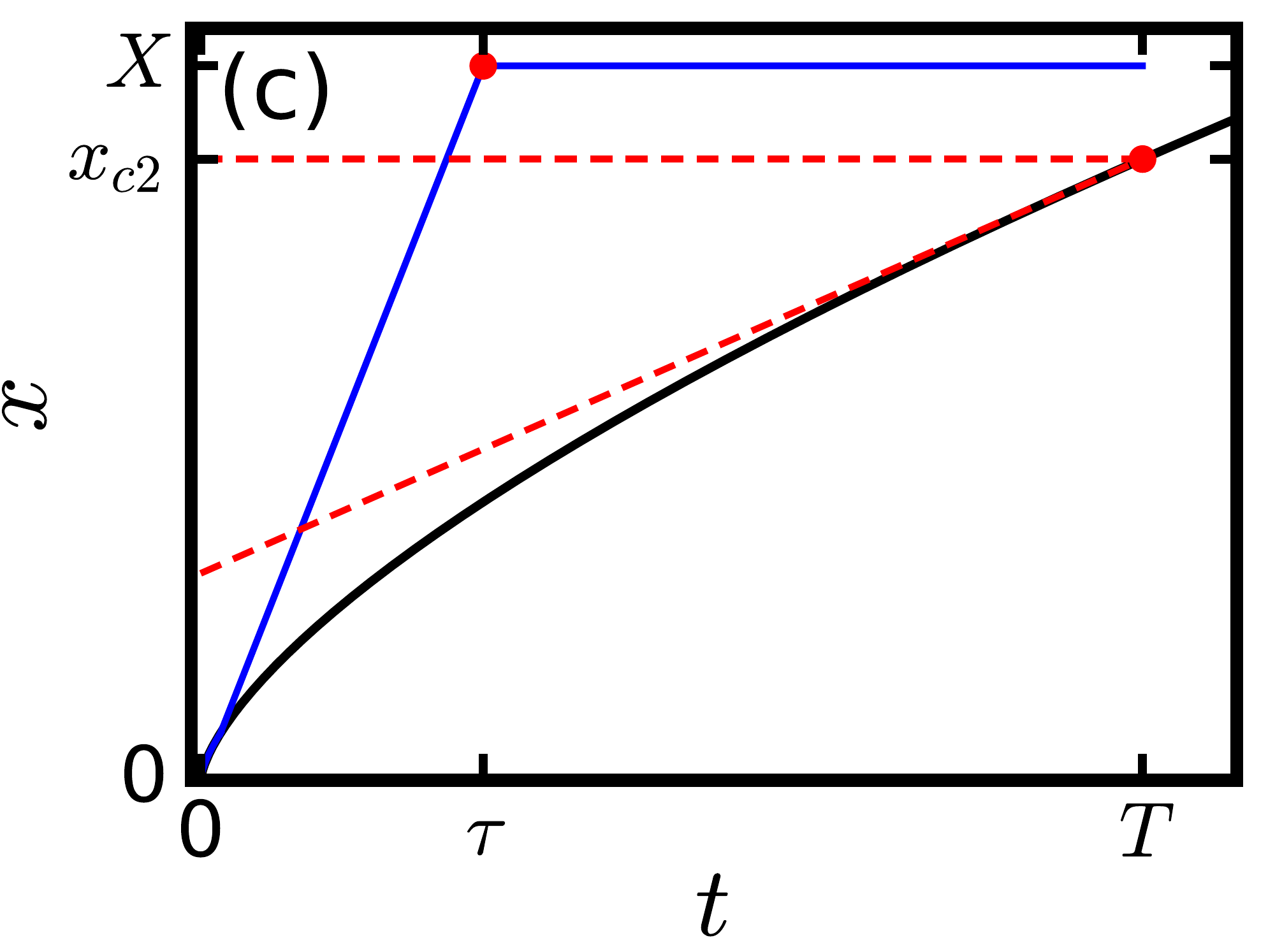}
\caption{The optimal path constrained on survival up to time $T$ and on the location $X$ at an earlier time $0 < \tau < T$ in the (a) subcritical (b) first supercritical and (c) second supercritical regimes for $x_{\text{w}}\left(t\right)\sim t^{2/3}$. The boundaries between the subcritical and first supercritical regimes and between the first and second supercritical regimes are dynamical phase transition lines of third and second order, respectively.}
\label{fig:intermediate_location}
\end{figure}

Motivated by Refs.~\cite{FS,SmithMeerson2019}, we  now ask an additional question. Given that the particle has not been absorbed until time $T$, what is the distribution $\mathcal{P}\left(X,\tau,T\right)$ of its location $X=x\left(t=\tau\right)$ at an earlier time $\tau$? %
As in Ref.~\citep{SmithMeerson2019}, we should first find the optimal path constrained on both nonabsorption, $x\left(t\right)> x_{\text{w}}\left(t\right)$ and on the value of $X$.
Let us again consider the wall functions $x_{\text{w}}\left(t\right)=Ct^{\gamma}$, for $1/2 < \gamma < 1$.
At long times, the distribution has the scaling form
\begin{equation}
\label{eq:intermediate_time_scaling}
-\ln\mathcal{P}\left(X,\tau,T\right)\simeq\frac{C^2T^{2\gamma-1}}{D}\mathcal{S}\left(\frac{X}{CT^{\gamma}},\frac{\tau}{T}\right)
\end{equation}
where the large-deviation function $\mathcal{S}$ is given by the difference between the actions~(\ref{Action}) evaluated on the optimal path and on the wall function \citep{SmithMeerson2019}. One should distinguish between three regimes. At subcritical $X$,
$$
C\tau^{\gamma}< X\le x_{\text{c}1}=CT^{\gamma}\left(1-\gamma+\frac{\gamma\tau}{T}\right),
$$
the optimal constrained path $x\left(t\right)$ is given by constructing two tangents from the point $\left(\tau,X\right)$ to the graph of the wall function $x_{\text{w}}\left(t\right)$, see Fig.~\ref{fig:intermediate_location} (a).
In the first supercritical regime,
$$
x_{\text{c}1}\le X\le x_{\text{c}2}=CT^{\gamma},
$$
the first segment of the optimal path, $0\le t\le\tau$, is given by the same tangent construction as in the subcritial regime, while the second segment  becomes of constant velocity:
\begin{equation}
x\left(t\right)=X+\frac{\left(t-\tau\right)\left(CT^{\gamma}-X\right)}{T-\tau},\qquad\tau\le t\le T,
\end{equation}
see Fig.~\ref{fig:intermediate_location} (b).
In the second supercritical regime, $X\ge x_{\text{c}2}$, the particle stops after reaching $x=X$,
see Fig.~\ref{fig:intermediate_location} (c).
The dynamical phase transition at $X = x_{\text{c}1}$ is third order (as in Ref.~\cite{SmithMeerson2019}), whereas the transition at $X = x_{\text{c}2}$ is second order.

One can calculate the action and the probability distribution \eqref{eq:intermediate_time_scaling}, by using Eqs.~\eqref{Action}, \eqref{sgammamore1} and the optimal paths that we have just found. We do not show here cumbersome formulas for the resulting large deviation function of $X$, 
but we present the near tail of the distribution:
\begin{equation}
\label{eq:moving_wall_near_tail}
-\ln\mathcal{P}(X,\tau)\simeq\frac{2\sqrt{2\gamma\left(1-\gamma\right)}
\,C^{\frac{1}{2}}\tau^{\frac{\gamma}{2}-1}}{3D}
\left(X-C\tau^{\gamma}\right)^{3/2},\qquad X-C\tau^{\gamma}\ll C\tau^{\gamma}.
\end{equation}
This tail is mostly contributed to by a small vicinity of the measurement time $\tau$, so it is independent of $T$. Because of this crucial property, the typical fluctuations of $X$,
\begin{equation}
X-C\tau^{\gamma}\sim
\left(\frac{C^{\frac{1}{2}}D}{\tau^{\frac{\gamma}{2}-1}}\right)^{2/3} ,
\end{equation}
which are generally beyond the accuracy of the OFM, should  obey the FS distribution \citep{FS}, which we already encountered in story 2. That is,
\begin{equation}
\label{eq:moving_wall_FS_distribution}
\mathcal{P}\left(X,\tau\right)\simeq\frac{\kappa\text{Ai}
\left[\kappa\left(X-C\tau^{\gamma}\right)+a_{1}\right]^{2}}
{\text{Ai}'\left(a_{1}\right)^{2}} ,\qquad \text{where} \quad \kappa=\left[\frac{-\ddot x_{\text{w}}\left(\tau\right)}{2D^{2}}\right]^{1/3}
=\left[\frac{\gamma\left(1-\gamma\right)\,C\tau^{\gamma-2}}{2D^{2}}\right]^{1/3}.
\end{equation}
As a check, we compared the right hand side of Eq.~(\ref{eq:moving_wall_near_tail}) with the expression in the exponent of the $\kappa\left(X-C\tau^{\gamma}\right)\gg1$ asymptotic of the FS distribution~(\ref{eq:moving_wall_FS_distribution}). They perfectly
agree.  

In the \emph{very far} tail of the distribution, $X\gg CT^{\gamma}$, the optimal path is simply
\begin{equation}
x\left(t\right)\simeq\begin{cases}
\frac{tX}{\tau}, & 0\le t\le\tau,\\
X, & \tau\le t\le T.
\end{cases}
\end{equation}
As a result, the very far tail is a simple free-particle Gaussian $-\ln\mathcal{P}\simeq X^{2}/\left(4D\tau\right),$ and the wall has no effect in the leading order, as to be expected.

\section{Discussion}

\label{disc}

We hope that these three short stories clearly demonstrated the advantages of geometrical optics for evaluating the statistics of Brownian motions, pushed into a large deviation regime by constraints.  One of the advantages of the geometrical optics is the knowledge of the optimal path of the Brownian particle, conditioned on a specified large deviation.  The optimal path language explains in a transparent way why strongly constrained Brownian motions often display dynamical phase transitions.  These come from geometric shadows:
either  in the configuration space as in stories 1 and 2, or in space-time as in story 3 (see also Refs. \cite{SmithMeerson2019,Meerson2019}).

The Ferrari-Spohn (FS) distribution features prominently in the regime of typical fluctuations in stories 2 and 3. While story 3 is very similar to the original context (escape from  a swinging wall) in which the FS distribution was first encountered \citep{FS}, the emergence of the FS distribution in story 2 is remarkable. We presented an argument where, exploiting a scale separation, one can replace one stochastic coordinate [in our case $x\left(t\right)$] by its deterministic optimal path counterpart. It would be very interesting to try and implement this argument in additional, more complicated multidimensional problems where a scale separation is present.  In any case, it appears that the FS distribution applies in more general settings than the swinging wall setting \citep{FS} in which it was originally observed.  The optimal path language gives a visual explanation of the universality of the FS distribution in terms of the locality of the constrained optimal path. Additional indications of the universality come from mathematical literature \cite{Ioffe2015,Ioffe2018a,Ioffe2018b,Ioffe2018c}.

The large-deviation (or rate) functions  $g$ in Eq.~(\ref{anyangle}), $s$ in Eq.~(\ref{eq:probability_scaling_Nechaev_problem}) and $\mathcal{S}$ in Eq.~(\ref{eq:intermediate_time_scaling}) are not affected by the boundary conditions on the wall: they are the same for absorbing and reflecting walls. The effect of the type of the boundary condition is much stronger in the regime of typical fluctuations.

The geometrical optics misses pre-exponential factors that can be interesting. A natural next step is to capture these factors by performing a saddle-point evaluation of the properly constrained path integral of the Brownian motion beyond the leading order.

Among potential important applications of geometrical optics of Brownian motion is the rate theory of biochemical processes.
In many biological systems, there is a very large number of ``searchers" (signal molecules, sperm cells, \textit{etc}.) which ``compete" for a single target cite (a cell surface receptor, an oocyte, \textit{etc}.). This huge redundancy is apparently exploited by nature in order to reduce the search time \cite{MR,Holcman}. As a result, the arrival of the first among the very many searchers to the target is unusually fast and can be analyzed by using geometrical optics \cite{Holcman}.

\section*{ACKNOWLEDGMENTS}

We thank T. Agranov,  A. Grosberg, P. Krapivsky, S. Nechaev and K. Polovnikov for useful discussions.
This research was supported by the Israel Science Foundation (grant No. 807/16).
N.R.S. was supported by the Clore Israel Foundation.



\begin{appendices}
\section{Mapping to the Ferrari-Spohn model for generalized parabolas}
\label{appendix:gen_parabola}

\renewcommand{\theequation}{A\arabic{equation}}
\setcounter{equation}{0}

In Ref.~\cite{SmithMeerson2019} we studied large deviations in the FS model over a time interval $|t|<T$  with general convex wall functions
\begin{equation}
\label{eq:x_wall_t}
x_{\text{w}}\left(t\right)=CT^{\gamma}g\left(t/T\right).
\end{equation}
We showed there that the distribution of $X = x(\tau)$ scales, in the large-deviation regime, as
\begin{equation}
\label{lnapp}
-\ln\mathcal{P}\left(X,\tau,T\right)\simeq\frac{C^{2}T^{2\gamma-1}}{D}\,s\left(\frac{X}{CT^{\gamma}},\frac{\tau}{T}\right).
\end{equation}
For rescaled wall functions
\begin{equation}
g\left(\frac{t}{T}\right)=1-\left|\frac{t}{T}\right|^{\lambda},
\end{equation}
with $\lambda>0$, we found the large-deviation function $s$ exactly. At $1<X\leq\lambda$ it is given by \citep{SmithMeerson2019}
\begin{equation}
\label{sapp}
s\left(1<X\leq\lambda\right)=\frac{\lambda^{2}\left(\lambda-1\right)}{2\lambda-1}\left(\frac{X-1}{\lambda-1}\right)^{\frac{2\lambda-1}{\lambda}}.
\end{equation}
This regime includes the regime of interest to us here, which is the near tail of the distribution $\Delta X=X-1\ll1$.
Using Eqs.~(\ref{lnapp}) and (\ref{sapp}), we obtain the near tail
\begin{equation}
\label{eq:near_tail_gen_parabola_FS_model}
-\ln\mathcal{P}\left(X,T\right)\simeq\frac{C^{2}T^{2\gamma-1}}{D}\frac{\lambda^{2}\left(\lambda-1\right)}{2\lambda-1}\left[\frac{\Delta X}{\left(\lambda-1\right)CT^{\gamma}}\right]^{\frac{2\lambda-1}{\lambda}}.
\end{equation}
In fact, Eq.~(\ref{eq:near_tail_gen_parabola_FS_model}) holds for all (convex) rescaled wall functions whose behavior in the vicinity of $t=\tau$ is
\begin{equation}
g\left(\frac{t}{T}\right)=g_{0}-\left|\frac{t-\tau}{T}\right|^{\lambda}+\dots,
\end{equation}
for arbitrary $\tau$ and arbitrary $g_0 > 0$. We now rewrite the effective wall function~(\ref{eq:gen_parabola_effective_wall}) in the form
\begin{equation}
y_{\text{w}}\left(t\right)=\alpha\mathcal{L}^{\lambda}\left(\frac{R}{\alpha\mathcal{L}^{\lambda}}-\left|\frac{t-\tau}{T}\right|^{\lambda}+\dots\right).
\end{equation}
This corresponds to Eq.~(\ref{eq:x_wall_t}) with $C=\alpha\mathcal{L}^{\lambda}$ and $\gamma=0$.
Plugging these parameters into Eq.~(\ref{eq:near_tail_gen_parabola_FS_model}), it is straightforward to see that the resulting distribution coincides exactly with our Eq.~(\ref{eq:near_tailgeneral}).

\end{appendices}

\end{document}